\documentclass[runningheads]{llncs}
\usepackage{graphicx}
\usepackage{tikz}
\usepackage{comment}
\usepackage{amsmath,amssymb} % define this before the line numbering.
\usepackage{color}

\usepackage{adjustbox}
\usepackage{multirow}
\usepackage{subfigure}
\usepackage{epsfig,epstopdf}
\usepackage{bbding}
\usepackage{cite}
\usepackage[colorlinks]{hyperref}

\usepackage{url}
\usepackage{array}

\newcommand{\ie}{i.e., }
\newcommand{\eg}{e.g., }
\newcommand{\etal}{\textit{et} \textit{al}.}

\newcolumntype{C}[1]{>{\centering\let\newline\\\arraybackslash\hspace{0pt}}m{#1}}

\begin{document}
% \linenumbers
\pagestyle{headings}
\mainmatter
\def\ECCVSubNumber{181}  % Insert your submission number here

\title{Texture Hallucination for Large-Factor Painting Super-Resolution} % Replace with your title

\begin{comment}
\titlerunning{ECCV-20 submission ID \ECCVSubNumber} 
\authorrunning{ECCV-20 submission ID \ECCVSubNumber} 
\author{Anonymous ECCV submission}
\institute{Paper ID \ECCVSubNumber}
\end{comment}
%******************

%\begin{comment}
\titlerunning{Texture Hallucination for Large-Factor Painting Super-Resolution}
\author{Yulun Zhang\inst{1} \and
Zhifei Zhang\inst{2} \and
Stephen DiVerdi\inst{2} \and\\
Zhaowen Wang\inst{2} \and
Jose Echevarria\inst{2} \and
Yun Fu\inst{1}}
\authorrunning{Y. Zhang et al.}
\institute{Northeastern University, USA \and
Adobe Research, USA
}
%\end{comment}
%******************
\maketitle
%\vspace{-4mm}
% at most 150 words
\begin{abstract}
We aim to super-resolve digital paintings, synthesizing realistic details from high-resolution reference painting materials for very large scaling factors (\eg 8$\times$, 16$\times$). However, previous single image super-resolution (SISR) methods would either lose textural details or introduce unpleasing artifacts. On the other hand, reference-based SR (Ref-SR) methods can transfer textures to some extent, but is still impractical to handle very large factors and keep fidelity with original input. To solve these problems, we propose an efficient high-resolution hallucination network for very large scaling factors with efficient network structure and feature transferring. To transfer more detailed textures, we design a wavelet texture loss, which helps to enhance more high-frequency components. At the same time, to reduce the smoothing effect brought by the image reconstruction loss, we further relax the reconstruction constraint with a degradation loss which ensures the consistency between downscaled super-resolution results and low-resolution inputs.
We also collected a high-resolution (\eg 4K resolution) painting dataset PaintHD by considering both physical size and image resolution. We demonstrate the effectiveness of our method with extensive experiments on PaintHD by comparing with SISR and Ref-SR state-of-the-art methods.
\end{abstract}

%\vspace{-3mm}
\section{Introduction}
%\vspace{-2mm}
Image super-resolution (SR) aims to reconstruct high-resolution (HR) output with details from its low-resolution (LR) counterpart. Super-resolution for digital painting images has important values in both culture and research aspects. Many historical masterpieces were damaged and their digital replications are in low-resolution (LR), low-quality due to technological limitation in old days. Recovery of their fine details is crucial for maintaining and protecting human heritage. It is also a valuable research problem for computer scientists to restore high-resolution (HR) painting due to the rich content and texture of paintings in varying scales. A straightforward way to solve this problem is to borrow some knowledge from natural image SR~\cite{dong2014learning,zhang2018image,zhang2019image}, which however is not enough. 

Super-resolving painting images is particularly challenging as vary large upscaling factors ($8\times$, $16\times$, or even larger) are required to recover the brush and canvas details of artworks, so that a viewer can fully appreciate the aesthetics as from the original painting. One state-of-the-art (SOTA) single image super-resolution (SISR) method RCAN~\cite{zhang2018image} can upscale input with large scaling factors with high PSNR values. But, it would suffer from over-smoothing artifacts, because most high-frequency components (\eg textures) have been lost in the input. It's hard to recover high-frequency information from LR input directly. Some reference-based SR (Ref-SR) methods try to transfer high-quality textures from another reference image. One SOTA Ref-SR method SRNTT~\cite{zhang2019image} matches features between input and reference. Then, feature swapping is conducted in a multi-level way. SRNTT performs well in the texture transfer. However, the results of SRNTT could be affected by the reference obviously. Also, it's hard for SRNTT to transfer high-quality textures when scaling factor becomes larger.      

Based on the analyses above, we try to transfer detailed textures from reference images and also tackle with large scaling factors. Fortunately, there is a big abundance of existing artworks scanned in high-resolution, which provide the references for the common texture details shared among most paintings. 
%Therefore, in this paper, we formulate a Ref-SR problem setting for digital paintings with large upscaling factors.

\begin{figure}[t]
\scriptsize
\centering
\newcommand{\wcell}{.23\columnwidth}
\begin{tabular}{C{\wcell}C{\wcell}C{\wcell}C{\wcell}}
HR & RCAN~\cite{zhang2018image} & SRNTT~\cite{zhang2019image} & Ours 
\end{tabular}
			
\includegraphics[width=.23\columnwidth]{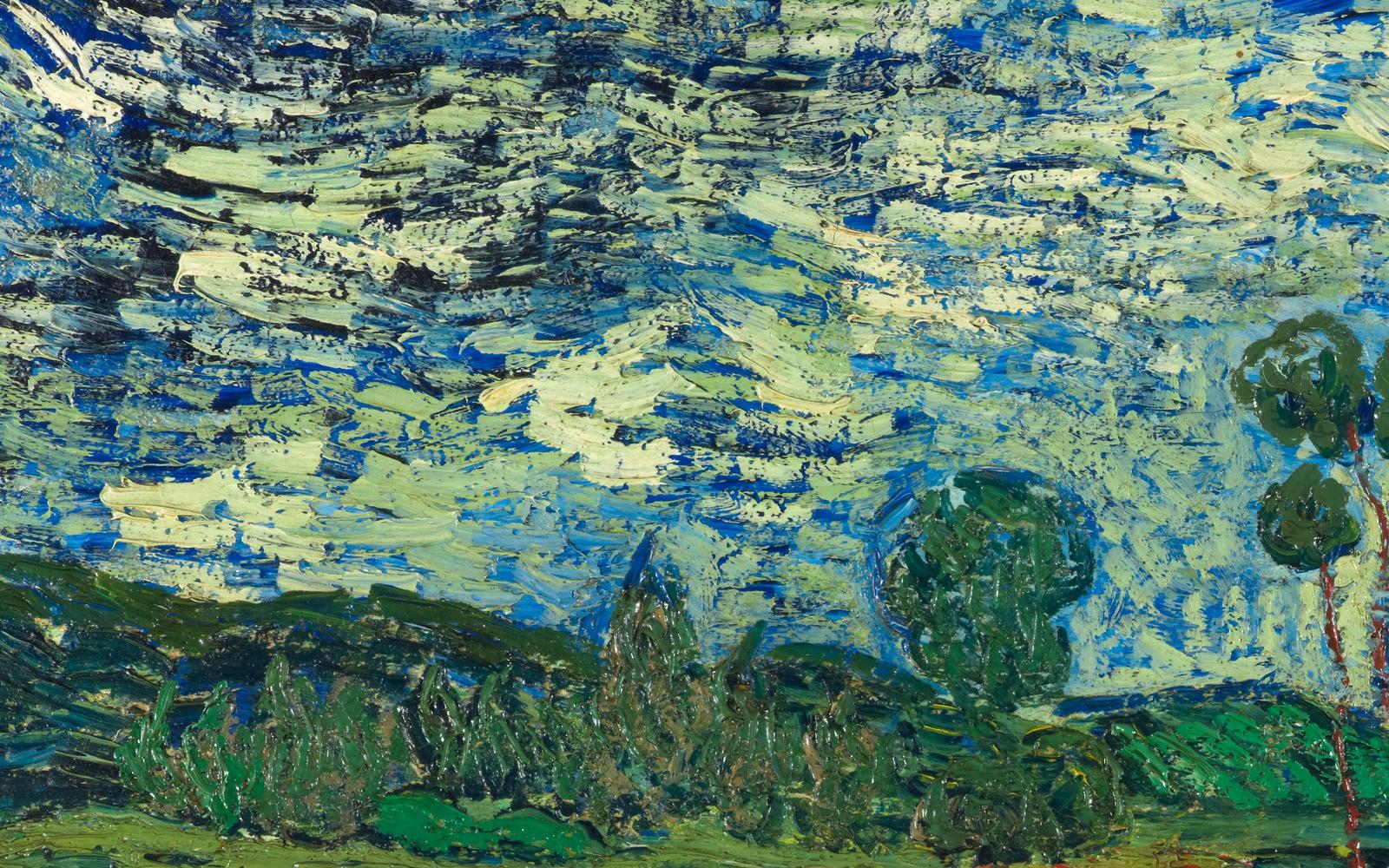}
\includegraphics[width=.23\columnwidth]{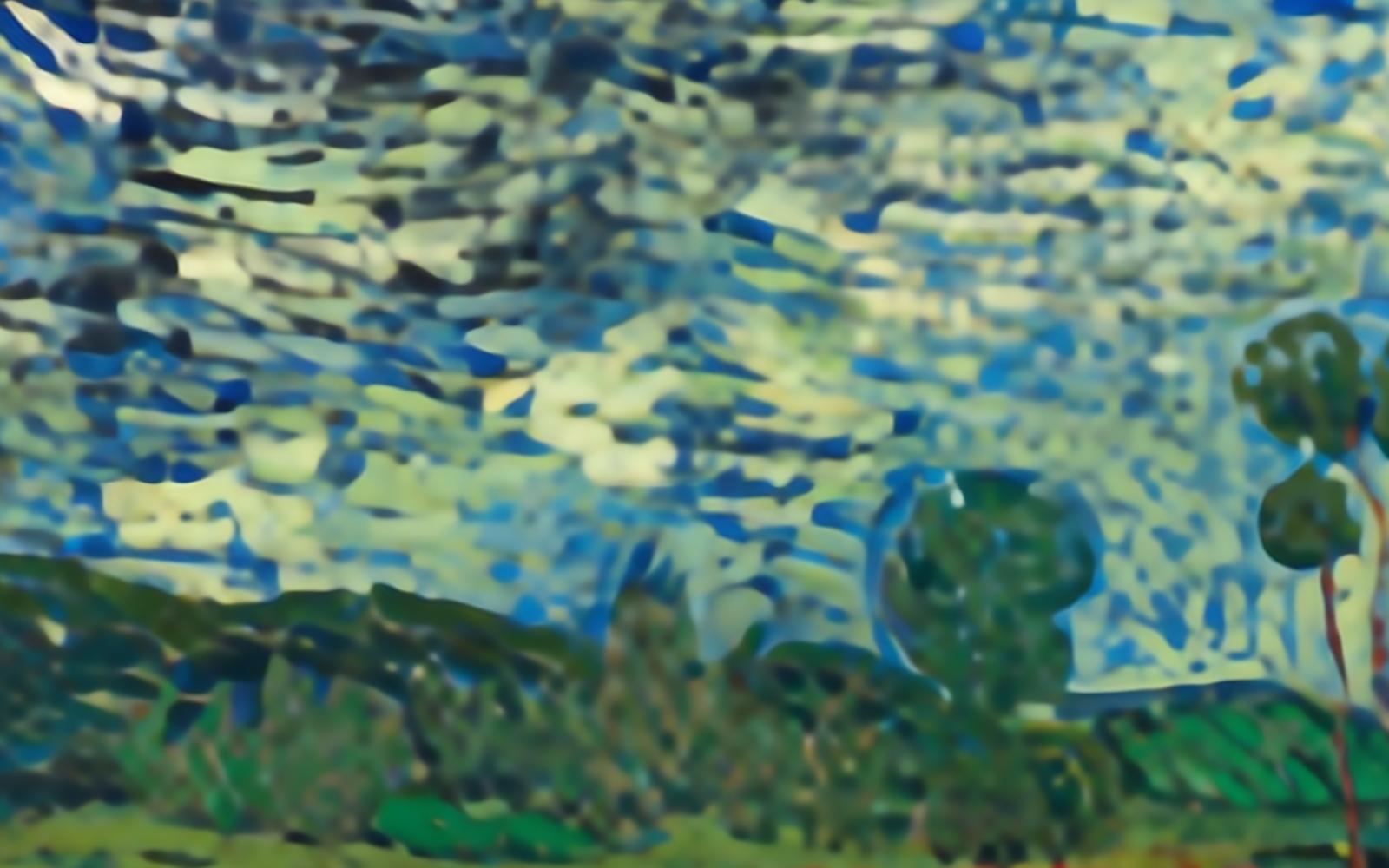}
\includegraphics[width=.23\columnwidth]{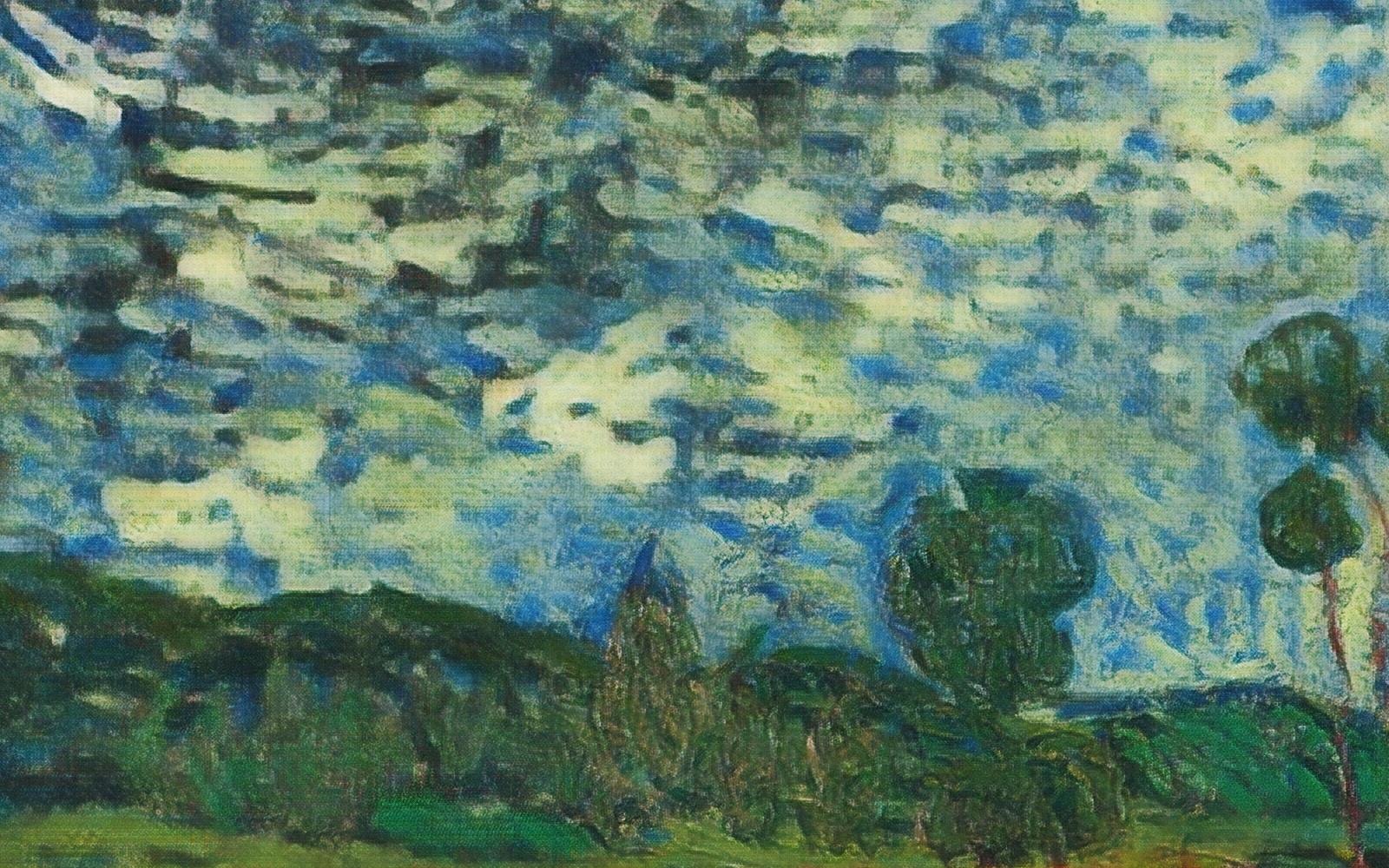}
\includegraphics[width=.23\columnwidth]{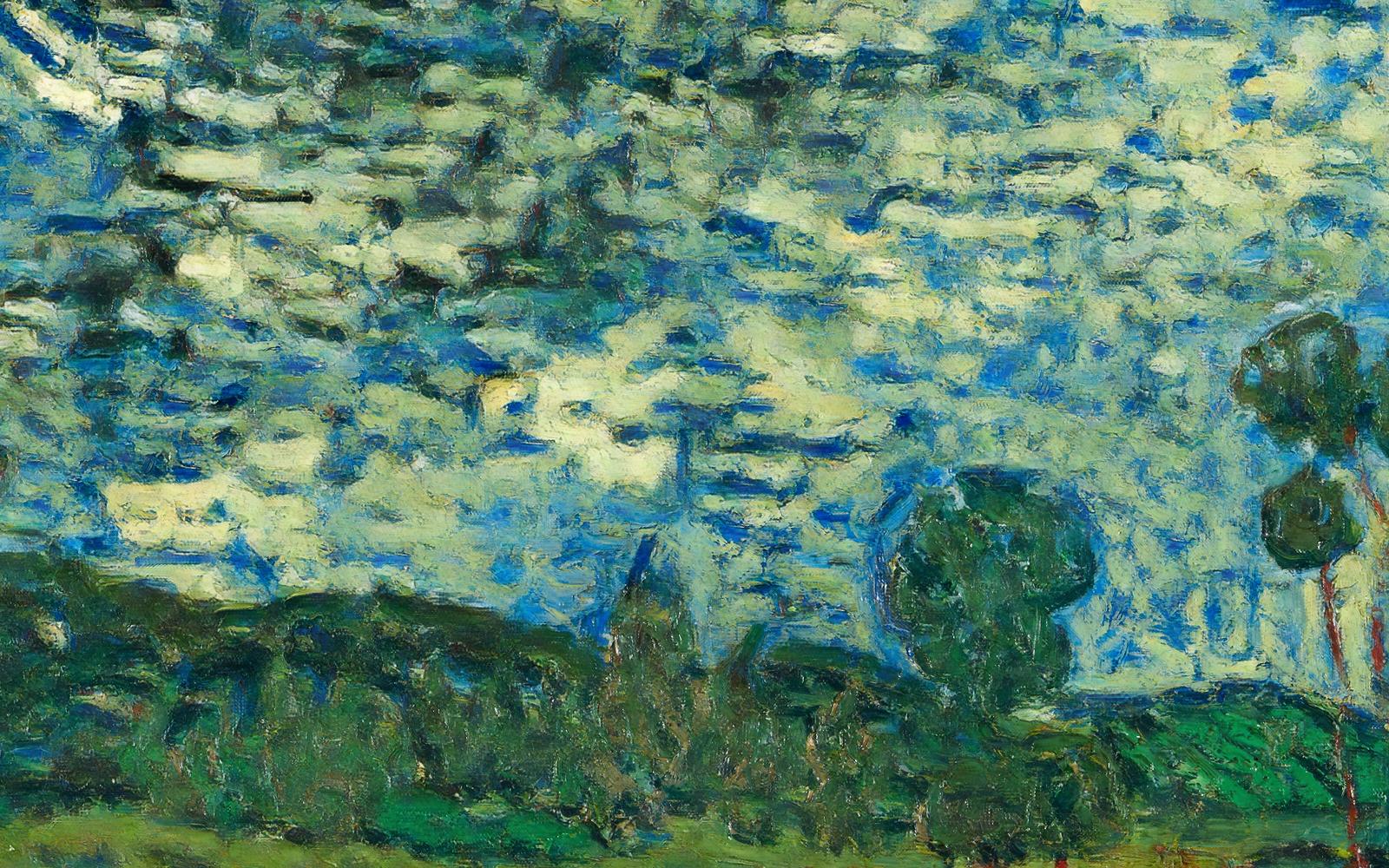}
						
\includegraphics[width=.23\columnwidth]{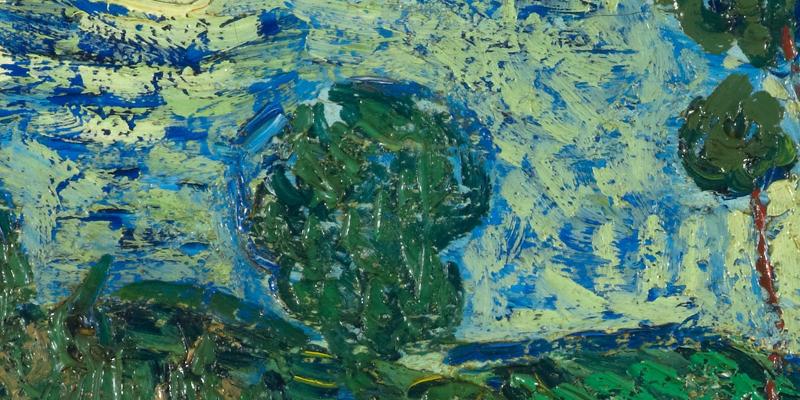}
\includegraphics[width=.23\columnwidth]{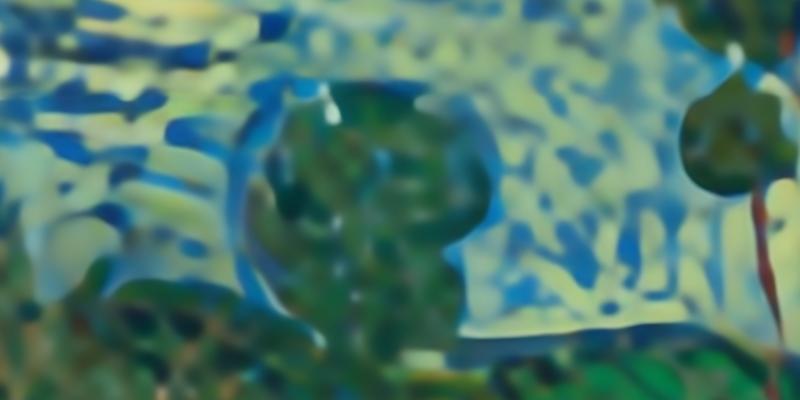}
\includegraphics[width=.23\columnwidth]{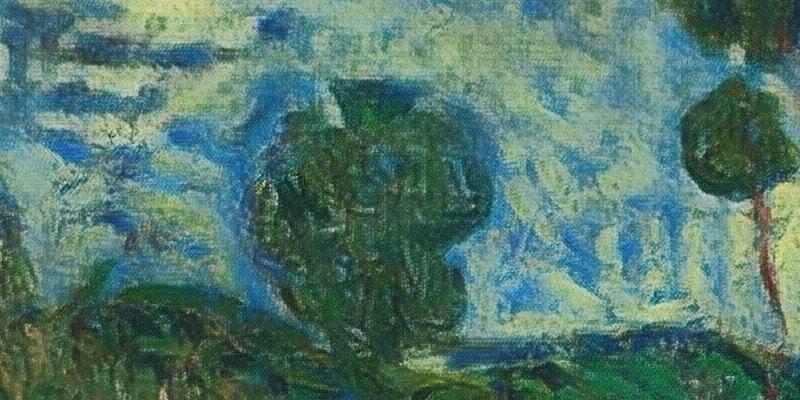}
\includegraphics[width=.23\columnwidth]{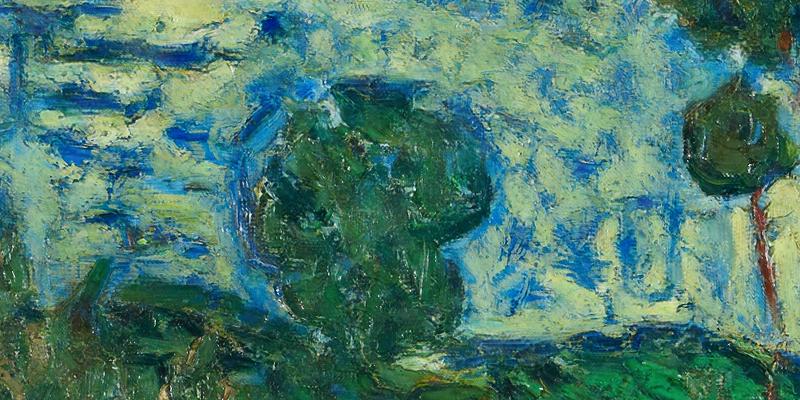}

%\vspace{-4mm}			
\caption{Visual comparisons for the scaling factor of 16$\times$ (the first row) and zoom-in patches (the second row). We compare with state-of-the-art SISR and Ref-SR methods.}
%(Please further zoom-in for better view)			
\label{fig:first_visual_results}
%\vspace{-4mm}
\end{figure}

To this end, we collect a large-scale high-quality dataset PaintHD for oil painting images with diverse contents and styles. We explore new deep network architectures with efficient texture transfer (\ie match feature in smaller scale and swap feature in fine scale) for large upscaling factors. We also design wavelet-based texture loss and degradation loss to achieve high perceptual quality and fidelity at the same time. The network architecture helps tackle large scaling factors better. The wavelet-based texture loss and degradation loss contribute to achieve better visual results. Our proposed method can hallucinate realistic details based on the given reference images, which is especially desired for large factor image upscaling. Compared to the previous SOTA SISR and Ref-SR methods, our proposed method achieves significantly improved quantitative (perceptual index (PI)~\cite{blau20182018}) and visual results, which are further verified in our human subjective evaluation (\ie user study). In Fig.~\ref{fig:first_visual_results}, we compare with other state-of-the-art methods for large scaling factor (\eg $16\times$). We can see our method can transfer more vivid and faithful textures. 

In summary, the main contributions of this work are:
\begin{itemize}
\item We proposed a reference-based image super-resolution framework for large upscaling factors (\eg $8\times$ and $16\times$) with novel training objectives. Specifically, we proposed wavelet texture loss and degradation loss, which allow to transfer more detailed and vivid textures.
\item We collected a new digital painting dataset PaintHD with high-quality images and detailed meta information, by considering both physical and resolution sizes. Such a high-resolution dataset is suitable for painting SR.
\item We achieved significantly improved quantitative and visual results over previous single image super-resolution (SISR) and reference based SR (Ref-SR) state-of-the-arts. A new technical direction is opened for Ref-SR with large upscaling factor on painting images.
\end{itemize}

%\vspace{-3mm}
\section{Related Work}
\label{sec:related_work}
%\vspace{-3mm}
Recent work on deep-learning-based methods for image SR~\cite{lim2017enhanced,sajjadi2017enhancenet,ledig2017photo,zhang2018image,zhang2018residual,zhang2019image} is clearly outperforming more traditional methods~\cite{freeman2000learning,chang2004super,sun2012super} in terms of either PSNR/SSIM or visual quality. Here, we focus on the former for conciseness. 
%\vspace{-3mm}
\subsection{Single Image Super-Resolution}
\label{subsec:sisr}
%\vspace{-3mm}
Single image super-resolution (SISR) recovers a high-resolution image directly from its low-resolution (LR) counterpart. The pioneering SRCNN proposed by Dong \etal~\cite{dong2014learning}, made the breakthrough of introducing deep learning to SISR, achieving superior performance than traditional methods. Inspired by this seminal work, many representative works~\cite{wang2015deep,kim2016accurate,kim2016deeply,dong2016accelerating,shi2016real,lim2017enhanced,zhang2018image} were proposed to further explore the potential of deep learning and have continuously raised the baseline performance of SISR. In SRCNN and follow-ups VDSR~\cite{kim2016accurate} and DRCN~\cite{kim2016deeply}, the input LR image is upscaled to the target size through interpolation before fed into the network for recovery of details. Later works demonstrated that extracting features from LR directly and learning the upscaling process would improve both quality and efficiency. For example, Dong \etal~\cite{dong2016accelerating} provide the LR image directly to the network and use a deconvolution for upscaling. Shi \etal~\cite{shi2016real} further speed up the upscaling process using sub-pixel convolutions, which became widely adopted in recent works. Current state-of-the-art performance is achieved by EDSR~\cite{lim2017enhanced} and RCAN~\cite{zhang2018image}. EDSR takes inspiration from ResNet~\cite{he2016deep}, using long-skip and sub-pix convolutions to achieve stronger edge and finer texture. RCAN introduced channel attention to learn high-frequency information. 

Once larger upscaling factors were achievable, \eg 4$\times$, 8$\times$, many empirical studies~\cite{ledig2017photo,sajjadi2017enhancenet,zhang2019image} demonstrated that the commonly used quality measurements PSNR and SSIM proved to be not representative of visual quality, \ie higher visual quality may result in lower PSNR; a fact first investigated by Johnson \etal~\cite{johnson2016perceptual} and Ledig \etal~\cite{ledig2017photo}. The former investigated perceptual loss using VGG~\cite{simonyan2014very}, while the later proposed SRGAN by introducing GAN~\cite{goodfellow2014generative} loss into SISR, which boosted significantly the visual quality compared to previous works. Based on SRGAN~\cite{ledig2017photo}, Sajjadi \etal~\cite{sajjadi2017enhancenet} further adopted texture loss to enhance textural reality. Along with higher visual quality, those GAN-based SR methods also introduce artifacts or new textures synthesized depending on the content, which would contribute to increased perceived fidelity.
% at the expense of reduced PSNR.

Although SISR has been studied for decades, it is still limited by its ill-posed nature, making it difficult to recover fine texture detail for upscaling factors of 8$\times$ or 16$\times$. So, most existing SISR methods are limited to a maximum of 4$\times$. Otherwise, they suffer serious degradation of quality. Works that attempted to achieve 8$\times$ upscaling, \eg LapSRN~\cite{lai2017deep} and RCAN~\cite{zhang2018image}, found visual quality would degrade quadratically with the increase of upscaling factor.
%\vspace{-3mm}
\subsection{Reference-based Super-Resolution}
%\vspace{-3mm}
Different from SISR, reference-based SR (Ref-SR) methods attempt to utilize self or external information to enhance the texture. Freeman \etal~\cite{freeman2002example} proposed the first work on Ref-SR, which replaced LR patches with fitting HR ones from a database/dictionary. \cite{freedman2011image,huang2015single} considered the input LR image itself as the database, from which references were extracted to enhance textures. These methods benefit the most from repeated patterns with perspective transformation. Light field imaging is an area of interest for Ref-SR, where HR references can be captured along the LR light field, just with a small offset. Thus, making easier to align the reference to the LR input, facilitating the transfer of high-frequency information in~\cite{boominathan2014improving,zheng2017combining}. CrossNet~\cite{zheng2018crossnet} took advantage of deep learning to align the input and reference by estimating the flow between them and achieved SOTA performance.

A more generic scenario for Ref-SR is to relax the constraints on references, \ie the references could present large spacial/color shift from the input. More extremely, references and inputs could contain unrelated content. Sun \etal~\cite{sun2012super} used global scene descriptors and internet-scale image databases to find similar scenes that provide ideal example textures. Yue \etal~\cite{yue2013landmark} proposed a similar idea, retrieving similar images from the web and performing global registration and local matching. Recent works~\cite{yang2018reference,zhang2019image} leveraged deep models and significantly improved Ref-SR performance, \eg visual quality and generalization capacity.

Our proposed method further extends the feasible scaling factor of previous Ref-SR methods from 4$\times$ to 16$\times$. More importantly, as oppose to the previous approach~\cite{zhang2019image}, which transfers the high-frequency information from reference as a style transfer task, we conduct texture transfer only in high-frequency band, which reduces the transfer effect on the low-frequency content.
%(and hence the distortion)
%\vspace{-3mm} 
\section{Approach}
%\vspace{-3mm}
We aim to hallucinate the SR image $I_{SR}$ for large scaling factor $s$ from its low-resolution (LR) input $I_{LR}$ and transfer highly detailed textures from high-resolution (HR) reference $I_{Ref}$. However, most previous Ref-SR methods~\cite{zheng2018crossnet,zhang2019image} could mainly handle relatively small scaling factors (\eg $\times4$). To achieve visually pleasing $I_{SR}$ with larger scaling factors, we firstly build a more compact network (see Fig.~\ref{fig:pipeline}) and then apply novel loss functions to the output. 
%\vspace{-3mm}
\subsection{Pipeline}
\begin{figure}[t]
\centering{
\includegraphics[width=.9\columnwidth]{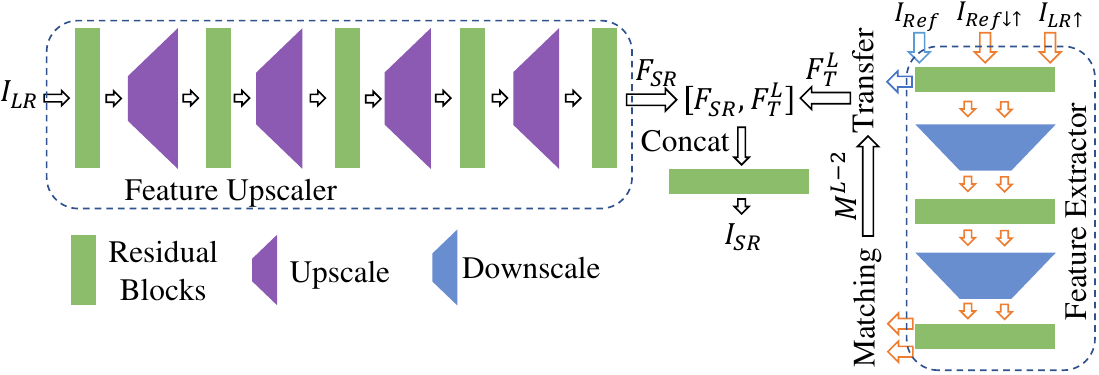}
}
%\vspace{-2mm}
\caption{The pipeline of our proposed method.}
\label{fig:pipeline}
%\vspace{-6mm}
\end{figure}

%\vspace{-3mm}
We first define $L$ levels according to scaling factor $s$, where $s=2^{L}$. Inspired by SRNTT~\cite{zhang2019image}, we conduct texture swapping in the feature space to transfer highly detailed textures to the output (Fig.~\ref{fig:pipeline}). The feature upscaler acts as the mainstream of upscaling the input LR image. Meanwhile, the reference feature that carries richer texture is extracted by the deep feature extractor. At the finest layer (largest scale) the reference feature is transferred to the output. 

As demonstrated in recent works~\cite{lee2015deeply,zhang2018image}, the batch normalization (BN) layers commonly used in deep models for stabilizing the training process turns to degrade the SR performance. Therefore, we avoid BN layer in our feature upscaling model.  
%However, we make some simplifications and improvements to handle larger scaling factors. Specifically, we first remove all the batch normalization (BN) layers in SRNTT. We find that to stabilize the training process.  
More importantly, the GPU memory usage is largely reduced, as the BN layer consumes similar amount of GPU memory as convolutional layer~\cite{lim2017enhanced}. %Second, different from SRNTT, which swaps features at multiple levels, we only swap features at the finest level $L$. This simplification saves extra GPU memory and running time.

To efficiently transfer high-frequency information from the reference, we swap features at the finest level $L$, where the reference features are swapped according to the local feature matching between the input and reference. Since patch matching is time-consuming, it is conducted at lower level (small spatial size), \ie we obtain feature matching information $M^{L-2}$ in level $L-2$ via
% narrow space
\setlength{\abovedisplayskip}{2pt}
\setlength{\belowdisplayskip}{2pt} 
\begin{align}
\begin{split}
\label{eq:M_L_2}
M^{L-2}=H^{L-2}_{M}\left ( \phi ^{L-2}\left ( I_{LR\uparrow} \right ),\phi ^{L-2}\left ( I_{Ref\downarrow\uparrow} \right ) \right ),
\end{split}
\end{align}
where $H^{L-2}_{M}(\cdot)$ denotes feature matching operation in level $L-2$. $\phi ^{L-2}(\cdot)$ is a neural feature extractor (\eg VGG19~\cite{simonyan2014very}) matching the same level. $I_{LR\uparrow}$ is upscaled by Bicubic interpolation with scaling factor $s$. To match the frequency band of $I_{LR\uparrow}$, we first downscale and then upscale it with scaling factor $s$. For each patch from $\phi ^{L-2}\left ( I_{LR\uparrow} \right )$, we could find its best matched patch from $\phi ^{L-2}\left ( I_{Ref\downarrow\uparrow} \right )$ with highest similarity.  

Then, using the matching information $M^{L-2}$, we transfer features at level $L$ and obtain the new feature $F^{L}_{T}$ via
\begin{align}
\begin{split}
\label{eq:F_swap}
F^{L}_{T}=H_{T}^{L}\left ( \phi ^{L}\left ( I_{Ref} \right ),M^{L-2} \right ),
\end{split}
\end{align}
where $H_{T}^{L}(\cdot)$ denotes feature transfer operation. $\phi ^{L}\left ( I_{Ref} \right )$ extracts neural feature from the high-resolution reference $I_{Ref}$ at level $L$.
	
On the other hand, we also extract deep feature from the LR input $I_{LR}$ and upscale it with scaling factor $s$. Let's denote the upscaled input feature as $F_{SR}$ and the operation as $H_{FSR}(\cdot)$, namely $F_{SR}=H_{FSR}(I_{LR})$. To introduce the transferred feature $F^{L}_{T}$ into the image hallucination, we fuse $F^{L}_{T}$ and $F_{SR}$ by using residual learning, and finally reconstruct the output $I_{SR}$. Such a process can be expressed as follows
\begin{align}
\begin{split}
\label{eq:I_SR}
I_{SR}=H_{Rec}\left ( H_{Res}\left ( \left [ F_{SR},F_{T}^{L} \right ] \right )+F_{SR} \right ),
\end{split}
\end{align}
where $[ F_{SR},F_{T}^{L} ]$ refers to channel-wise concatenation, $H_{Res}(\cdot)$ denotes several residual blocks, and $H_{Rec}(\cdot)$ denotes a reconstruction layer. 
	
We can already achieve super-resolved results with larger scaling factors by using the above simplifications and improvements. The ablation study in Section~\ref{subsec:ablation_study} would demonstrate the effectiveness of the simplified pipeline. However, we still aim to transfer highly-detailed texture from reference even in such challenging cases (\ie very large scaling factors). To achieve this goal, we further propose wavelet texture and degradation losses.
%\vspace{-3mm}
\subsection{Wavelet Texture Loss}
%\vspace{-3mm}
\textbf{Motivation.} Textures are mainly composed of high-frequency components. LR images contain less high-frequency components, when the scaling factor goes larger. If we apply the loss functions (including texture loss) on the color image space, it's still hard to recover or transfer more high-frequency ones. However, if we pay more attention to the high-frequency components and relax the reconstruction of color image space, such an issue could be alleviated better. Specifically, we aim to transfer as many textures as possible from reference by applying texture loss on the high-frequency components. Wavelet is a proper way to decompose the signal into different bands with different frequency levels. 

\textbf{Haar wavelet.} Inspired by the excellent WCT$^2$~\cite{yoo2019photorealistic}, where a wavelet-corrected transfer was proposed, we firstly apply Haar wavelet to obtain different components. The Haar wavelet transformation has four kernels, $\left \{ LL^{T},LH^{T},HL^{T},HH^{T} \right \}$, where $L^T$ and $H^T$ denote the low and high pass filters,
\begin{align}
\begin{split}
\label{eq:wavelet}
L^{T}=\frac{1}{\sqrt{2}}\begin{bmatrix}
1 & 1
\end{bmatrix},\ 
H^{T}=\frac{1}{\sqrt{2}}\begin{bmatrix}
-1 & 1
\end{bmatrix}.
\end{split}
\end{align}
	
As a result, such a wavelet operation would split the signal into four channels, capturing low-frequency and high-frequency components. We denote the extraction operations for these four channels as $H_{W}^{LL}(\cdot)$, $H_{W}^{LH}(\cdot)$, $H_{W}^{HL}(\cdot)$, and $H_{W}^{HH}(\cdot)$ respectively. Then, we aim to pay more attention to the recovery of high-frequency components with the usage of wavelet texture loss.   
	
\textbf{Wavelet texture loss.} As investigated in WCT$^2$~\cite{yoo2019photorealistic}, in Haar wavelet, the low-pass filter can extract smooth surface and parts of texture and high-pass filters capture higher frequency components (\eg horizontal, vertical, and diagonal edge like textures). 
	
Ideally, it's a wise choice to apply texture loss on each channel split by Haar wavelet. However, as we calculate texture loss in different scales, such a choice would suffer from very heavy GPU memory usage and running time. Moreover, as it's very difficult for the network to transfer highly detailed texture with very large scaling factors, focusing on the reconstruction of more desired parts would be a better choice. Consequently, we propose a wavelet texture loss with $HH$ kernel and formulate it as follows      
\begin{align}
\begin{split}
\label{eq:loss_wave_tex}
\mathcal{L}_{tex}=\sum_{l}\lambda _{l}\left \| Gr\left ( \phi ^{l}\left ( H_{W}^{HH}\left ( I_{SR} \right ) \right ) \right )- Gr\left ( F_{T}^{l} \right ) \right \|_{F},
\end{split}
\end{align}
where $H_{W}^{HH}(\cdot)$ extracts high-frequency component from the upscaled output $I_{SR}$ with $HH$ kernel. $F_{T}^{l}$ is the transferred feature in feature map space of $\phi ^{l}$. $Gr(\cdot)$ calculates the Gram matrix for each level $l$, where $\lambda_{l}$ is the corresponding normalization weight. $\left \| \cdot\right \| _{F}$ denotes Frobenius norm.
	
As shown in Eq.~\eqref{eq:loss_wave_tex}, we mainly focus on the texture reconstruction of higher frequency components, which would transfer more textures with somehow creative ability. Then, we further relax the reconstruction constraint by proposing a degradation loss.

\begin{figure}[tpb]
\centering
\includegraphics[width=.5\columnwidth]{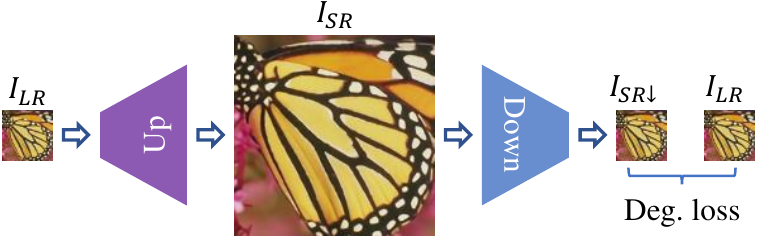}
%\vspace{-2mm}
\caption{The illustration of our proposed degradation loss. We try to minimize the degradation loss $\mathcal{L}_{deg}$ between the downscaled output $I_{SR\downarrow}$ and the original input $I_{LR}$.}
\label{fig:loss_deg}
%\vspace{-4mm}
\end{figure}
%\vspace{-3mm}
\subsection{Degradation Loss}
\label{subsec:loss_deg}
%\vspace{-3mm}
\textbf{Motivation.} Most previous single image SR methods (\eg RCAN~\cite{zhang2018image}) mainly concentrate on minimizing the loss between the upscaled image $I_{SR}$ and ground truth $I_{GT}$. For small scaling factors (\eg $\times$2), those methods would achieve excellent results with very high PSNR values. However, when the scaling factor goes very large (\eg $16\times$), the results of those methods would suffer from heavy smoothing artifacts and lack favorable textures (see Fig.~\ref{fig:first_visual_results}). On the other hand, as we try to transfer textures to the results as many as possible, emphasizing on the overall reconstruction in the upscaled image may also smooth some transferred textures. To alleviate such texture oversmoothing artifacts, we turn to additionally introduce the LR input $I_{LR}$ into network optimization.         
	
\textbf{Degradation loss.} Different from image SR, which is more challenging to obtain favorable results, image downscaling could be relatively easier. It's possible to learn a degradation network $H_{D}$, that maps the HR image to an LR one. We train such a network by using HR ground truth $I_{GT}$ as input and try to minimize the loss between its output $H_{D}(I_{GT})$ and the LR counterpart $I_{LR}$.
	
With the degradation network $H_{D}$, we are able to mimic the degradation process from $I_{GT}$ to $I_{LR}$, which can be a many-to-one case. Namely, there exists many upscaled images corresponding to the original LR image $I_{LR}$, which helps to relax the constraints on the reconstruction. To make use of this property, we try to narrow the gap between the downscaled output $I_{SR\downarrow}$ and the original LR input $I_{LR}$. As shown in Fig.~\ref{fig:loss_deg}, we formulate it as a degradation loss    
\begin{align}
\begin{split}
\label{eq:loss_deg}
\mathcal{L}_{deg}=\left \| I_{SR\downarrow} -I_{LR} \right \|_1=\left \| H_{D}\left ( I_{SR} \right ) -I_{LR} \right \|_1,
\end{split}
\end{align}
where $I_{SR\downarrow}$ denotes the downscaled image from $I_{SR}$ with scaling factor $s$ and $\left \| \cdot\right \| _{1}$ denotes $\ell_{1}$-norm. With the proposed loss functions, we further give details about the implementation.
%\vspace{-3mm}
%\subsection{Differences with SRNTT}
%\label{subsec:diff_srntt}
%\vspace{-3mm}
%As our work is developed from SRNTT~\cite{zhang2019image}, we would like to highlight two main differences between SRNTT and ours. The first one is the design of the pipeline. Specifically, we remove all the BN layers, which helps reduce GPU memory usage and also improve performance (it's investigated in EDSR~\cite{lim2017enhanced}). SRNTT conducts feature swap in multi-level, which may achieve higher PSNR/SSIM values, but hurt the perceptual quality. Instead, our method only conducts feature swap in the fine scale, which further saves running time and transfers textures more easily. The comparison between SRNTT and ours with $\mathcal{L}_{rec}$ only demonstrates the effectiveness of our pipeline. More details are given in Section~\ref{subsec:ablation_study}. The third one is that new loss functions $\mathcal{L}_{tex}$ and $\mathcal{L}_{deg}$ are used in our method. Such loss functions help to transfer more textures from high-quality reference, resulting in better perceptual quality. Compared with SRNTT, our method achieves better quantitative performance (\eg perceptual index (PI) and user study scores) and visual output (\eg more faithful textures to the ground truth).    
%\vspace{-3mm}
\subsection{Implementation Details}
\label{subsec:implementation}
%\vspace{-3mm}
\textbf{Loss functions.} We also adopt another three common loss functions~\cite{johnson2016perceptual,ledig2017photo,sajjadi2017enhancenet,zhang2019image}: reconstruction ($\mathcal{L}_{rec}$), perceptual ($\mathcal{L}_{per}$), and adversarial ($\mathcal{L}_{adv}$) losses. We briefly introduce them as follows.
\begin{align}
\begin{split}
\label{eq:loss_rec}
\mathcal{L}_{rec}=\left \| I_{SR}-I_{GT}\right \| _{1},
\end{split}
\end{align}
\begin{align}
\begin{split}
\label{eq:loss_per}
\mathcal{L}_{per}=\frac{1}{N_{5,1}}\sum_{i=1}^{N_{5,1}}\left\|\phi ^{5,1}_{i}\left ( I_{SR} \right ) -\phi ^{5,1}_{i}\left ( I_{GT} \right ) \right \|_{F},
\end{split}
\end{align}
where $\phi ^{5,1}$ extracts $N_{5,1}$ feature maps from 1-$st$ convolutional layer before 5-$th$ max-pooling layer of the VGG-19~\cite{simonyan2014very} network. $\phi ^{5,1}_{i}$ is the $i$-$th$ feature map.   
	
We also adopt WGAN-GP~\cite{gulrajani2017improved} for adversarial training~\cite{goodfellow2014generative}, which can be expressed as follows
\begin{align}
\begin{split}
\label{eq:loss_minG_maxD}
\min_{G}\max_{D}\mathbb{E}_{I_{GT}\sim\mathbb{P}_{r} }\left [ D(I_{GT}) \right ]-\mathbb{E}_{I_{SR}\sim\mathbb{P}_{g} }\left [ D\left ( I_{SR} \right ) \right ],
\end{split}
\end{align}
where $G$ and $D$ denote generator and discriminator respectively, and $I_{SR}=G(I_{LR})$. $\mathbb{P}_{r}$ and $\mathbb{P}_{g}$ represent data and model distributions. For simplicity, here, we mainly focus on the adversarial loss for generator and show it as follows
\begin{align}
\begin{split}
\mathcal{L}_{adv}=-\mathbb{E}_{I_{SR}\sim\mathbb{P}_{g} }\left [ D\left ( I_{SR} \right ) \right ].
\end{split}
\label{eq:loss_adv}
\end{align}
	
\textbf{Training.} The weights for $\mathcal{L}_{rec}$, $\mathcal{L}_{tex}$, $\mathcal{L}_{deg}$, $\mathcal{L}_{per}$, and $\mathcal{L}_{adv}$ are 1, 10$^{-4}$, 1, 10$^{-4}$, and 10$^{-6}$ respectively. To stabilize the training process, we pre-train the network for 2 epochs with $\mathcal{L}_{rec}$ and $\mathcal{L}_{tex}$. Then, all the losses are applied to train another 20 epochs. We implement our model with TensorFlow and apply Adam optimizer~\cite{kingma2014adam} with learning rate 10$^{-4}$.

\begin{figure}[tpb]
\scriptsize
\centering
\subfigure[\scriptsize{HR}]{\includegraphics[width=.19\columnwidth]{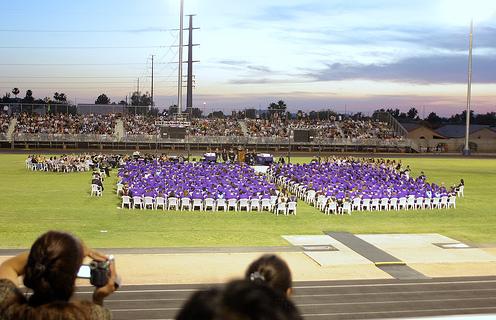}}
\subfigure[\scriptsize{Reference}]{\includegraphics[width=.19\columnwidth]{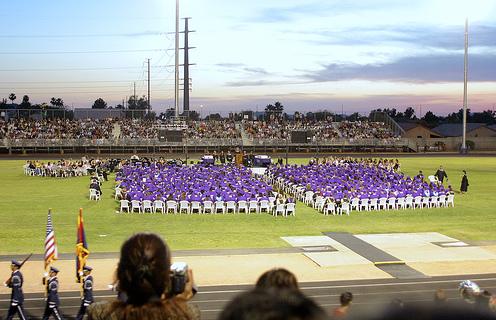}}
%\subfigure[Bicubic]{\includegraphics[width=.24\columnwidth]{figs_jpg/cmp/8x/001_0/Bicubic.jpg}}
\subfigure[\scriptsize{RCAN}]{\includegraphics[width=.19\columnwidth]{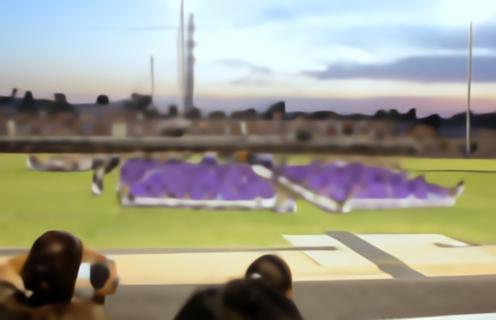}}
\subfigure[\scriptsize{SRNTT}]{\includegraphics[width=.19\columnwidth]{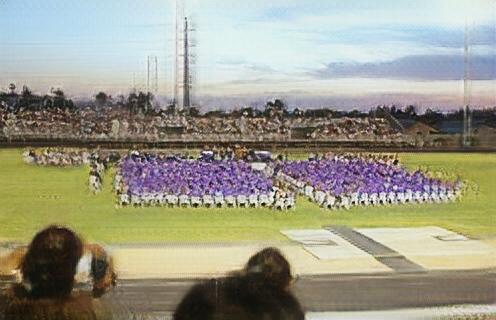}}
\subfigure[\scriptsize{Ours}]{\includegraphics[width=.19\columnwidth]{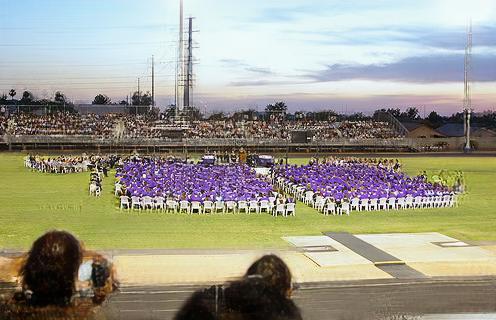}}
%\vspace{-4mm}
\caption{Visual results ($8\times$) of RCAN~\cite{zhang2018image}, SRNTT~\cite{zhang2019image}, and our method on CUFED5. Our result is visually  more pleasing than others, and generates plausible texture details.}
\label{fig:SR_x8_CUFED5}
%\vspace{-3mm}
\end{figure}
	
\begin{figure}[tpb]
\centering
\includegraphics[width=\columnwidth]{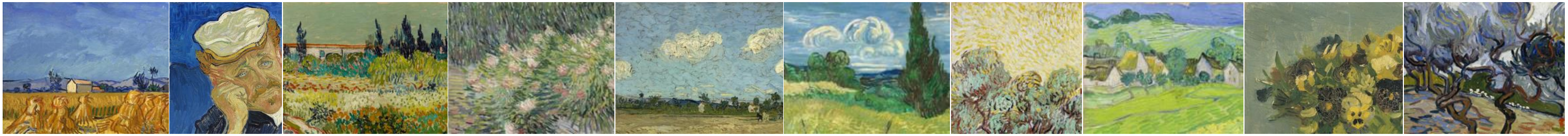}
%\vspace{-8mm}
\caption{Examples from our collected PaintHD dataset.}
\label{fig:painting_samples}
%\vspace{-4mm}
\end{figure}

%\vspace{-4mm}
\section{Dataset}
\label{sec:dataset}
%\vspace{-3mm}
For large upscaling factors, \eg 8$\times$ and 16$\times$, input images with small size, \eg 30$\times$30, but with rich texture in its originally HR counterpart will significantly increase the arbitrariness/smoothness for texture recovery because fewer pixels result in looser content constraints on the texture recovery. Existing datasets for Ref-SR are unsuitable for such large upscaling factors (see Fig.~\ref{fig:SR_x8_CUFED5}). Therefore, we collect a new dataset of high-resolution painting images that carry rich and diverse stroke and canvas texture. 
	
The new dataset, named PaintHD, is sourced from the Google Art Project~\cite{googleart}, which is a collection of very large zoom-able images. In total, we collected over 13,600 images, some of which achieve gigapixel. Intuitively, an image with more pixels does not necessarily present finer texture since the physical size of the corresponding painting may be large as well. To measure the richness of texture, the physical size of paintings is considered to calculate pixel per inch (PPI) for each image. Finally, we construct the training set consisting of 2,000 images and the testing set of 100 images with relatively higher PPI. Fig.~\ref{fig:painting_samples} shows some examples of PaintHD, which contains abundant textures. 
	
%We follow the same way described in SRNTT~\cite{zhang2019image} to collect input-reference image pairs, resulting in pairs.
	
%\zhifei{How to get input-ref pairs? How many pairs in total?}
	
To further evaluate the generalization capacity of the proposed method, we also test on the CUFED5~\cite{zhang2019image} dataset, which is designed specifically for Ref-SR validation. There are 126 groups of samples. Each group consists of one HR image and four references with different levels of similarity to the HR image. For simplicity, we adopt the most similar reference for each HR image to construct the testing pairs. The images in CUFED5 are of much lower resolution, \eg 500$\times$300, as compared to the proposed PaintHD dataset.

%\vspace{-4mm}
\section{Experimental Results}
%\vspace{-3mm}
%We conduct extensive experiments to validate the contributions of each component in our method. We demonstrate the effectiveness of our method by comparing with other state-of-the-art methods quantitatively and visually. We also provide more details, results, and analyses in supplementary material.
%\footnote{\href{http://yulunzhang.com/papers/PaintingSR\_supp\_arXiv.pdf}{http://yulunzhang.com/papers/PaintingSR\_supp\_arXiv.pdf}}.
%\vspace{-4mm}
\subsection{Ablation Study}
\label{subsec:ablation_study}
\begin{figure}[t]
%\newlength-4mm
%\setlength{-4mm}{-0.4cm}
\scriptsize
\centering
\begin{tabular}{cc}
% % one row
%\hspace{-0.4cm}
%\begin{adjustbox}{valign=t}
%\begin{tabular}{c}
%\includegraphics[width=0.2035\textwidth]{figs_jpg/ablation/pipeline/8x/ours_crop0.jpg}
%\\
%Kodak24: kodim24
%\end{tabular}
%\end{adjustbox}
%\hspace{-0.46cm}
\subfigure[8$\times$]{
\begin{adjustbox}{valign=t}
\begin{tabular}{cccccc}
\includegraphics[width=0.23\textwidth]{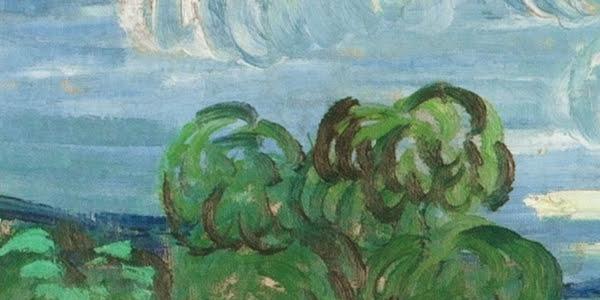} \hspace{-1mm} &
\includegraphics[width=0.23\textwidth]{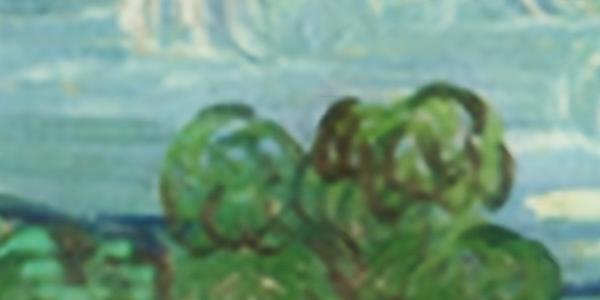} \hspace{-1mm} 
\\
HR \hspace{-4mm} &
SRNTT-$\mathcal{L}_{rec}$ \hspace{-4mm}
\\
\includegraphics[width=0.23\textwidth]{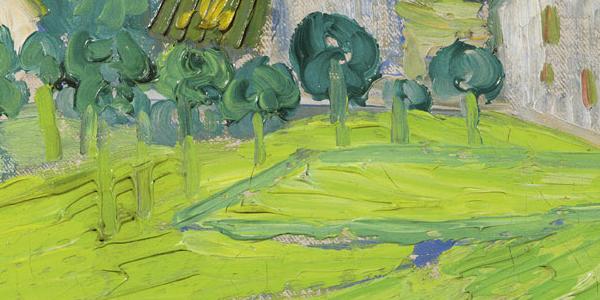} \hspace{-1mm} &
\includegraphics[width=0.23\textwidth]{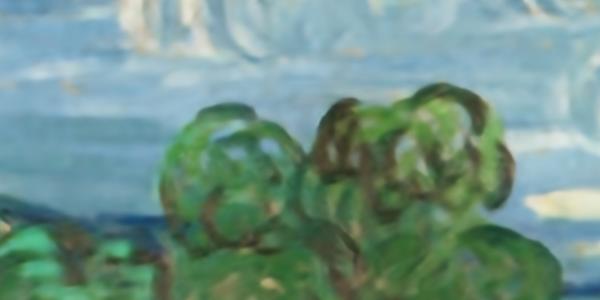} \hspace{-1mm}  
\\ 
Ref \hspace{-4mm} &
Ours-$\mathcal{L}_{rec}$ \hspace{-4mm}
\\
\end{tabular}
\end{adjustbox}
\label{fig:ablation_pipeline_8x}
}

\subfigure[16$\times$]{
\begin{adjustbox}{valign=t}
\begin{tabular}{cccccc}
\includegraphics[width=0.23\textwidth]{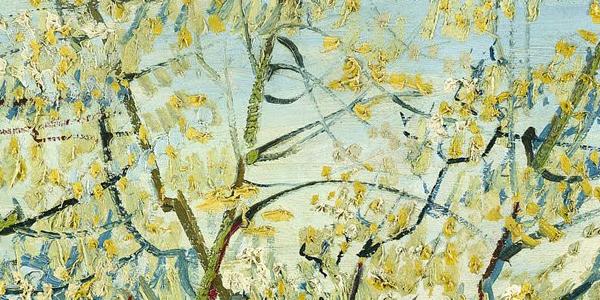} \hspace{-1mm} &
\includegraphics[width=0.23\textwidth]{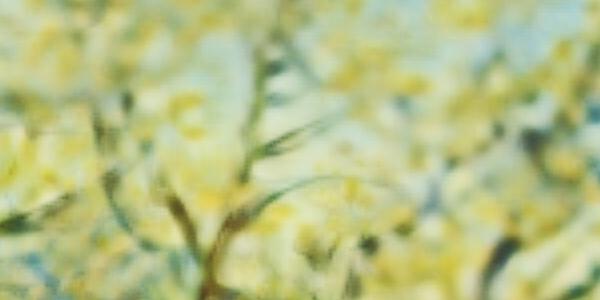} \hspace{-1mm} 
\\
HR \hspace{-4mm} &
SRNTT-$\mathcal{L}_{rec}$ \hspace{-4mm}
\\
\includegraphics[width=0.23\textwidth]{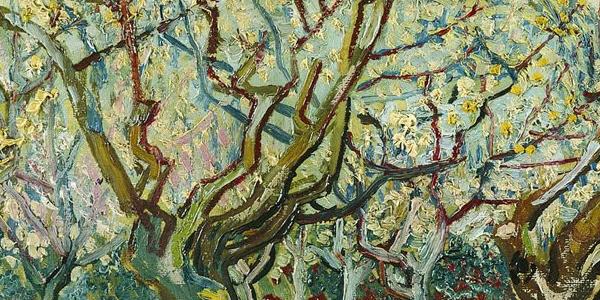} \hspace{-1mm} &
\includegraphics[width=0.23\textwidth]{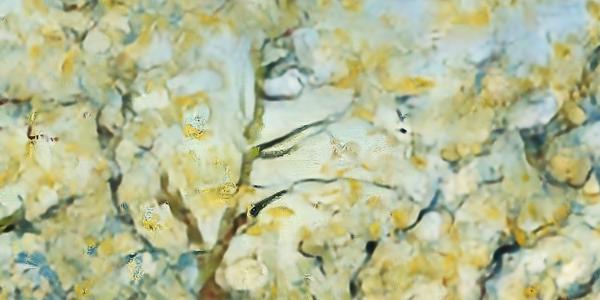} \hspace{-1mm}  
\\ 
Ref \hspace{-4mm} &
Ours-$\mathcal{L}_{rec}$ \hspace{-4mm}
\\
\end{tabular}
\end{adjustbox}
\label{fig:ablation_pipeline_16x}
}

\end{tabular}
%\vspace{-6mm}
\caption{Visual comparisons between SRNTT and ours by using $\mathcal{L}_{rec}$ only}
\label{fig:ablation_pipeline}
%\vspace{-4mm}
\end{figure}

%\vspace{-3mm}
%and also support our analyses about the difference with SRNTT~\cite{zhang2019image} in Section~\ref{subsec:diff_srntt}.
\textbf{Effect of Our Pipeline.} We firstly try to demonstrate the effectiveness of our simplified pipeline. We re-train SRNTT and our model by using PaintHD and reconstruction loss $\mathcal{L}_{rec}$ only with scaling factors 8$\times$ and 16$\times$. We show visual comparisons about 8$\times$ in Fig.~\ref{fig:ablation_pipeline_8x}. We can see the color of the background by our method is more faithful to the ground truth. Furthermore, our method achieves sharper result than that of SRNTT. Such a observation can be much clearer, when the scaling factor becomes 16$\times$ (\eg see Fig.~\ref{fig:ablation_pipeline_16x}). As a result, our method transfers more textures and achieve shaper results. We also provide quantitative results about `SRNTT-$\mathcal{L}_{rec}$' and `Ours--$\mathcal{L}_{rec}$' in Table~\ref{tab:results_psnr_ssim}, where we'll give more details and analyses. In summary, these comparisons demonstrate the effectiveness of our simplified pipeline.

\begin{figure}[t]
\scriptsize
\centering
\subfigure[\scriptsize{HR}]{\includegraphics[height=14mm]{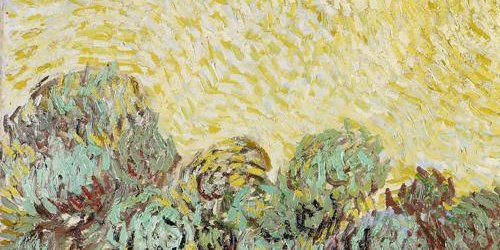}\label{subfig:hr}}
\subfigure[\scriptsize{Reference}]{\includegraphics[height=14mm]{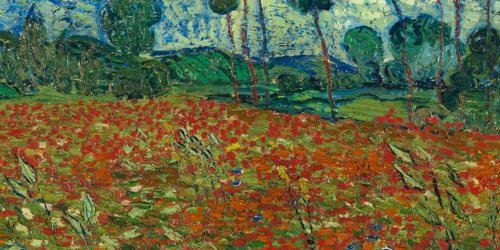}\label{subfig:ref}}
\subfigure[\scriptsize{w/o Wavelet}]{\includegraphics[height=14mm]{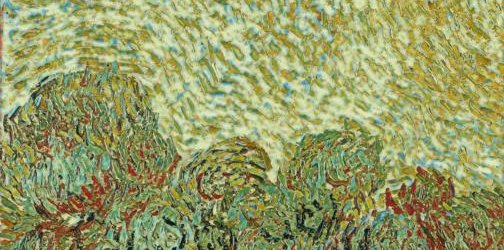}\label{subfig:wo_wavelet}}
\subfigure[\scriptsize{w/ Wavelet}]{\includegraphics[height=14mm]{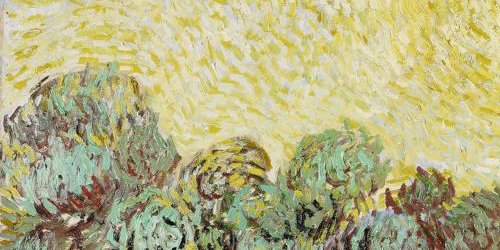}\label{subfig:w_wavelet}}
%\vspace{-4mm}
\caption{Comparison of super-resolved results with and without wavelet}
\label{fig:hh}
%\vspace{-6mm}
\end{figure}

\textbf{Effect of Wavelet Texture Loss.} The wavelet texture loss is imposed on the high-frequency band of the feature maps, rather than directly applying on raw features like SRNTT~\cite{zhang2019image} and traditional style transfer~\cite{gatys2016image}. Comparison between the wavelet texture loss and tradition texture loss is illustrated in Fig.~\ref{fig:hh}. To highlight the difference, weights on texture losses during training are increased by 100 times as compared to the default setting in Section~\ref{subsec:implementation}. Let's compare Figs.~\ref{subfig:wo_wavelet} and \ref{subfig:w_wavelet}, the result without wavelet is significantly affected by the texture/color from the reference (Fig.~\ref{subfig:ref}), lost identity to the input content. By contrast, the result with wavelet still preserves similar texture and color to the ground truth (Fig.~\ref{subfig:hr}).

\begin{figure}[t]
\scriptsize
\centering
\subfigure[\scriptsize{HR}]{\includegraphics[height=14mm]{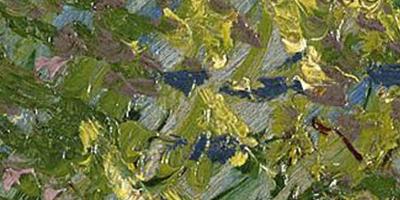}}
\subfigure[\scriptsize{Reference}]{\includegraphics[height=14mm]{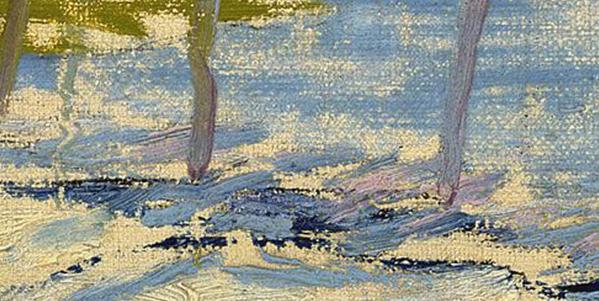}}
\subfigure[\scriptsize{w/o $\mathcal{L}_{deg}$}]{\includegraphics[height=14mm]{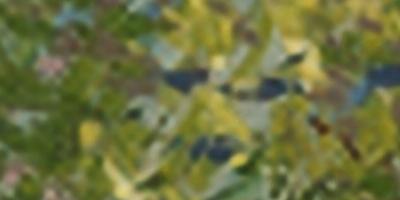}\label{fig:ablation_deg_loss_nodeg}}
\subfigure[\scriptsize{w/ $\mathcal{L}_{deg}$}]{\includegraphics[height=14mm]{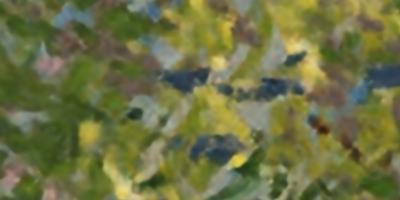}\label{fig:ablation_deg_loss_deg}}
%\vspace{-5mm}
\caption{Comparison of super-resolved results (8$\times$) with and without degradation loss}
\label{fig:ablation_deg_loss}
%\vspace{-6mm}
\end{figure}

\textbf{Effect of Degradation Loss.} To demonstrate the effectiveness of our proposed degradation loss $\mathcal{L}_{deg}$, we train one of our models with $\mathcal{L}_{rec}$ only and another same model with $\mathcal{L}_{rec}$ and $\mathcal{L}_{deg}$ with scaling factor $8\times$. We show the visual comparison in Fig.~\ref{fig:ablation_deg_loss}, where we can see result with $\mathcal{L}_{rec}$ only would suffer from some blurring artifacts (see Fig.~\ref{fig:ablation_deg_loss_nodeg}). While, in Fig.~\ref{fig:ablation_deg_loss_deg}, $\mathcal{L}_{deg}$ helps suppress such artifacts to some degree. This is mainly because the degradation loss $\mathcal{L}_{deg}$ alleviates the training difficulty in the ill-posed image SR problem. Such observations not only demonstrate the effectiveness of $\mathcal{L}_{deg}$, but also are consistent with our analyses in Section~\ref{subsec:loss_deg}.

%\vspace{-4mm}
\subsection{Quantitative Results}
\label{subsec:psnr_ssim}
%\vspace{-3mm}	
We compare our method with state-of-the-art SISR and Ref-SR methods. The SISR methods are EDSR~\cite{lim2017enhanced}, RCAN~\cite{zhang2018image}, and SRGAN~\cite{ledig2017photo}, where RCAN~\cite{zhang2018image} achieved state-of-the-art performance in terms of PSNR (dB). Due to limited space, we only introduce the state-of-the-art Ref-SR method SRNTT~\cite{zhang2019image} for comparison. However, most of those methods are not originally designed for very large scaling factors. Here, to make them suitable for $8\times$ and $16\times$ SR, we adopt them with some modifications. In $8\times$ case, we use RCAN~\cite{zhang2018image} to first upscale the input $I_{LR}$ by $2\times$. The upscaled intermediate result would be the input for EDSR and SRGAN, which then upscale the result by $4\times$. Analogically, in $16\times$ case, we use RCAN to first upscale $I_{LR}$ by $4\times$. The intermediate result would be def into RCAN, EDSR, and SRGAN, which further upscale it by $4\times$. For SRNTT and our method, we would directly upscale the input by $8\times$ or $16\times$. SRNTT is re-trained with our PaintHD training data by its authors.  

%\footnote{We use implementations from \\
%EDSR: \href{https://github.com/thstkdgus35/EDSR-PyTorch}{https://github.com/thstkdgus35/EDSR-PyTorch}\\
%RCAN: \href{https://github.com/yulunzhang/RCAN}{https://github.com/yulunzhang/RCAN}\\
%SRGAN: \href{https://github.com/tensorlayer/srgan}{https://github.com/tensorlayer/srgan}\\
%SRNTT: \href{https://github.com/ZZUTK/SRNTT}{https://github.com/ZZUTK/SRNTT}}

We not only compute the pixel-wise difference with PSNR and SSIM~\cite{wang2004image}, but also evaluate perceptual quality with perceptual index (PI)~\cite{blau20182018} by considering Ma's score~\cite{ma2017learning} and NIQE~\cite{mittal2012making}. Specifically, PI = 0.5((10 - Ma) + NIQE). Lower PI value reflects better perceptual quality. We show quantitative results in Table~\ref{tab:results_psnr_ssim}, where we have some interesting and thought-provoking observations.

\begin{table}[t]
\scriptsize
%\footnotesize
%\small
%\normalsize
\centering
\begin{center}
%%\vspace{-3mm}
\caption{Quantitative results (PSNR/SSIM/PI) of different SR methods for $8\times$ and $16\times$ on two datasets: CUFED5~\cite{zhang2019image} and our collected PaintHD. The methods are grouped into two categories: SISR (top group) and Ref-SR (bottom). We highlight the best results for each case. `Ours-$\mathcal{L}_{rec}$' denotes our method by using only $\mathcal{L}_{rec}$}
\label{tab:results_psnr_ssim}
%\vspace{-2mm}
%\begin{tabular*}{84mm}{@{\extracolsep{-0.97mm}}|c|c|c|c|c|c|c|c|c|c|c|c|c|c|c|c|c|}
\begin{tabular}{|c|c|c|c|c|c|c|c|c|c|c|c|c|c|c|c|c|c|}
\hline
Data &  \multicolumn{2}{c|}{CUFED5} &  \multicolumn{2}{c|}{PaintHD}    
\\
\hline
Scale &  $8\times$ &  $16\times$ & $8\times$ & $16\times$    
\\
\hline
\hline    
{Bicubic }
& 21.63/0.572/9.445 & 19.75/0.509/10.855 & 23.73/0.432/9.235 & 22.33/0.384/11.017
\\
%\hline
{EDSR}
& 23.02/0.653/7.098 & 20.70/0.548/8.249 & 24.42/0.477/7.648 & 22.90/0.405/8.943
\\
%\hline
{RCAN}
& \textcolor{red}{23.37}/\textcolor{red}{0.666}/6.722 & \textcolor{red}{20.71}/\textcolor{red}{0.548}/8.188 & \textcolor{red}{24.43}/\textcolor{red}{0.478}/7.448 & \textcolor{red}{22.91}/\textcolor{red}{0.406}/8.918
\\
%\hline
{SRGAN}
& 22.93/0.642/5.714 & 20.54/0.537/7.367 & 24.21/0.466/7.154 & 22.75/0.396/7.955
\\
\hline
\hline
{SRNTT-$\mathcal{L}_{rec}$}
& 22.34/0.612/7.234 & 20.17/0.528/8.373 & 23.96/0.449/7.992 & 22.47/0.391/8.464 
\\
{SRNTT}
& 21.08/0.548/2.502 & 19.09/0.418/2.956 & 22.90/0.377/3.856 & 21.48/0.307/4.314
\\

%\hline
{Ours-$\mathcal{L}_{rec}$}
& 22.40/0.635/4.520 & 19.71/0.526/5.298 & 24.02/0.461/5.253 & 22.13/0.375/5.815
\\
%\hline
{Ours}
& 20.36/0.541/\textcolor{red}{2.339} & 18.51/0.442/\textcolor{red}{2.499} & 22.49/0.361/\textcolor{red}{3.670} & 20.69/0.259/\textcolor{red}{4.131}
\\
\hline
\end{tabular}
\end{center}
%\vspace{-6mm}
\end{table}

\begin{figure}[t]
%\newlength-4mm
%\setlength{-4mm}{-0.4cm}
\scriptsize
\centering
\begin{tabular}{cc}
% % one row
%\hspace{-0.4cm}
%\begin{adjustbox}{valign=t}
%\begin{tabular}{c}
%\includegraphics[width=0.2035\textwidth]{figs_jpg/ablation/pipeline/8x/ours_crop0.jpg}
%\\
%Kodak24: kodim24
%\end{tabular}
%\end{adjustbox}
%\hspace{-0.46cm}
\begin{adjustbox}{valign=t}
\begin{tabular}{cccccc}
\includegraphics[width=0.24\textwidth]{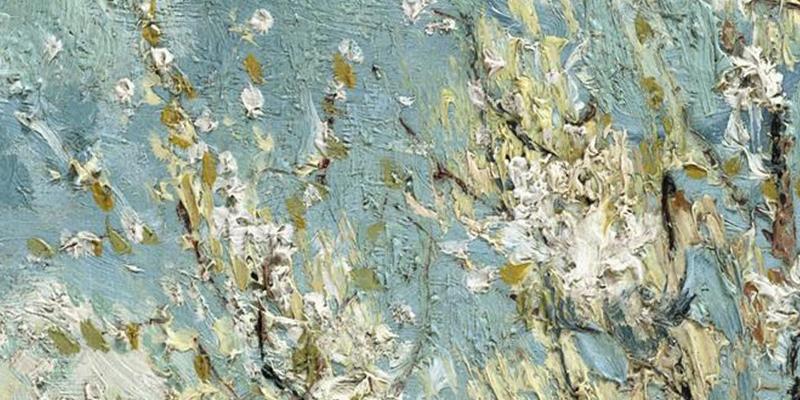} \hspace{-1mm} &
\includegraphics[width=0.24\textwidth]{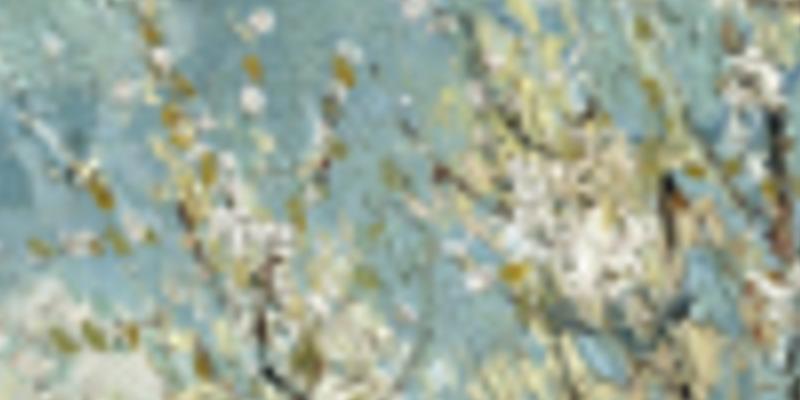} \hspace{-1mm} &
\includegraphics[width=0.24\textwidth]{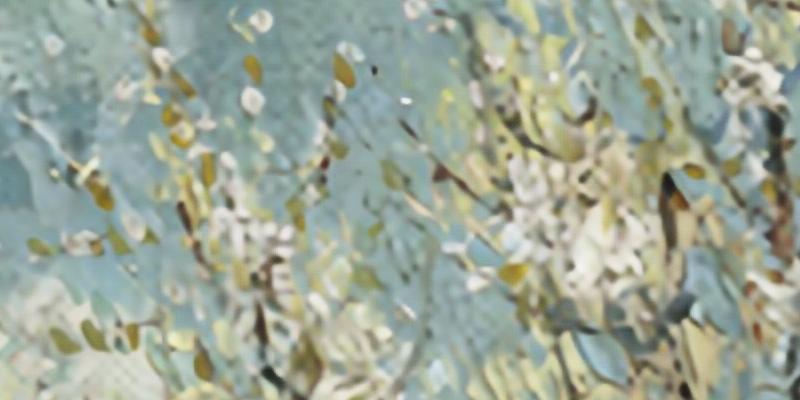} \hspace{-1mm} &
\includegraphics[width=0.24\textwidth]{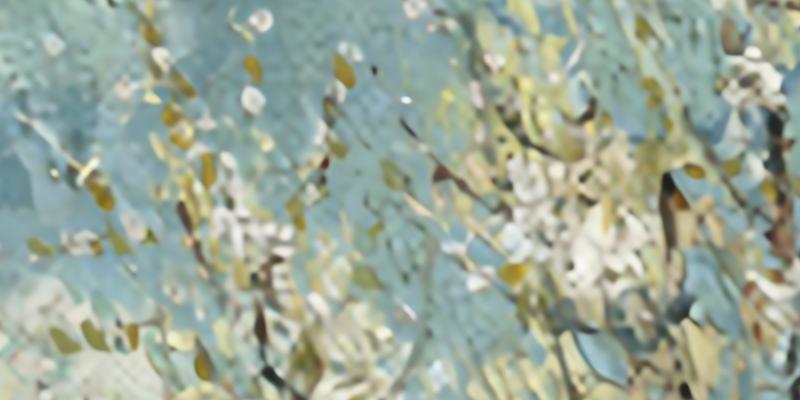} \hspace{-1mm} 
\\
HR \hspace{-4mm} &
Bicubic \hspace{-4mm} &
SRGAN~\cite{ledig2017photo} \hspace{-4mm} &
EDSR~\cite{lim2017enhanced} \hspace{-4mm}
\\
\includegraphics[width=0.24\textwidth]{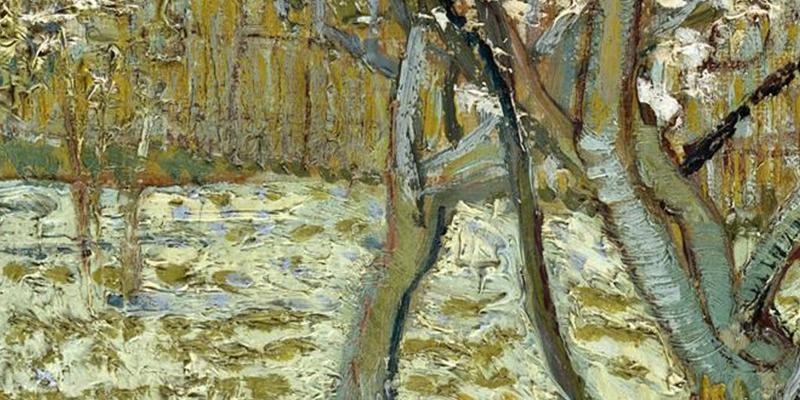} \hspace{-1mm} &
\includegraphics[width=0.24\textwidth]{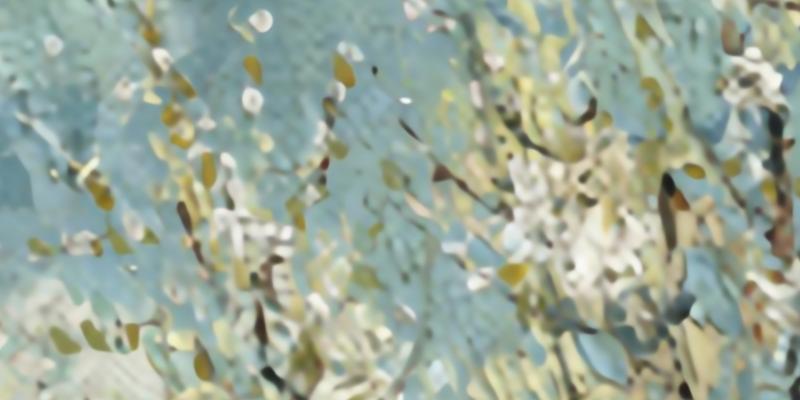} \hspace{-1mm} &
\includegraphics[width=0.24\textwidth]{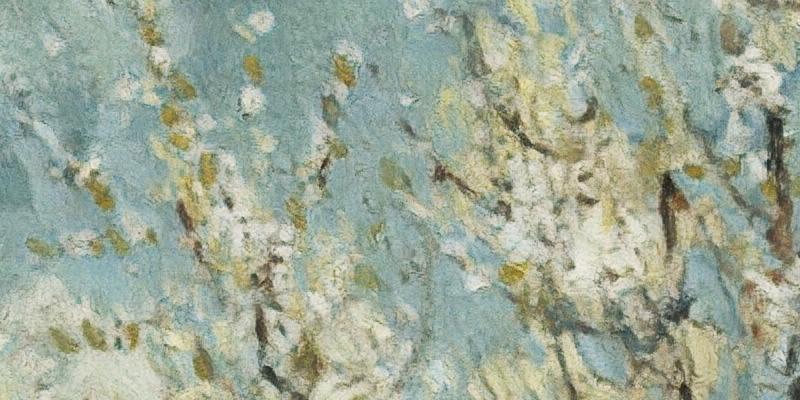} \hspace{-1mm} &
\includegraphics[width=0.24\textwidth]{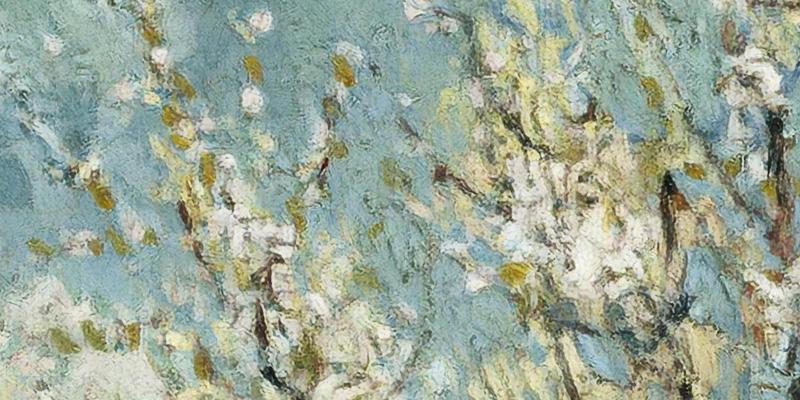} \hspace{-1mm}  
\\ 
Reference \hspace{-4mm} &
RCAN~\cite{zhang2018image} \hspace{-4mm} &
SRNTT~\cite{zhang2019image} \hspace{-4mm} &
Ours \hspace{-4mm}
\\
\end{tabular}
\end{adjustbox}

\\
\begin{adjustbox}{valign=t}
\begin{tabular}{cccccc}
\includegraphics[width=0.24\textwidth]{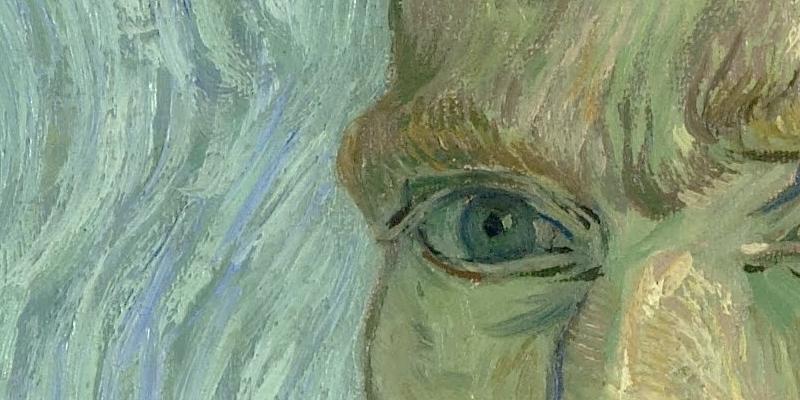} \hspace{-1mm} &
\includegraphics[width=0.24\textwidth]{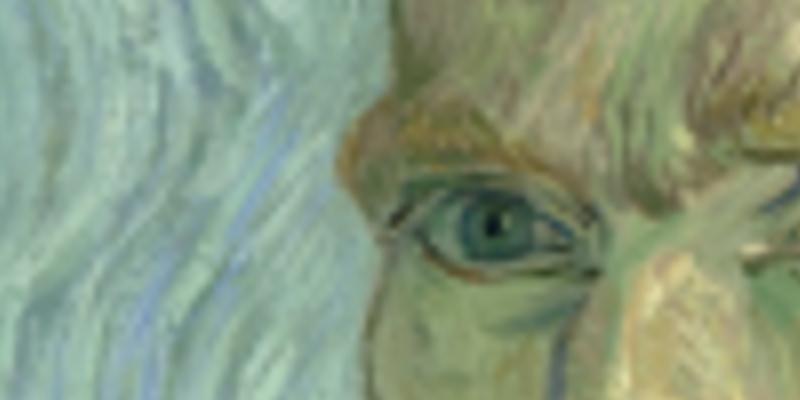} \hspace{-1mm} &
\includegraphics[width=0.24\textwidth]{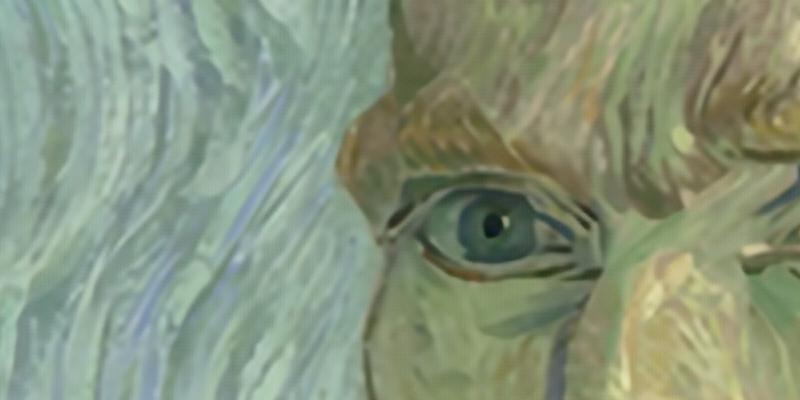} \hspace{-1mm} &
\includegraphics[width=0.24\textwidth]{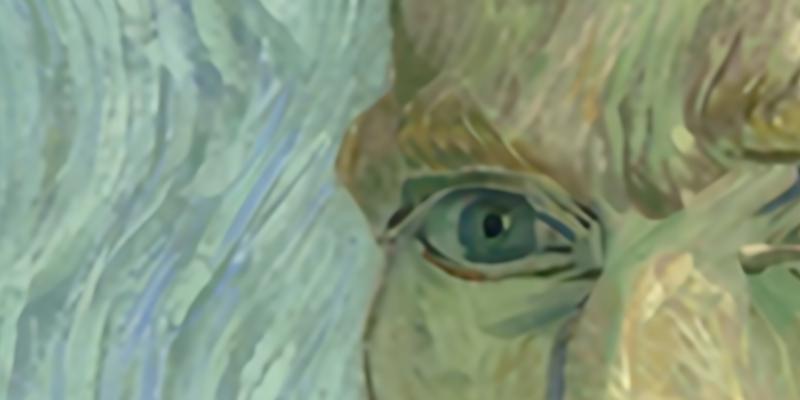} \hspace{-1mm} 
\\
HR \hspace{-4mm} &
Bicubic \hspace{-4mm} &
SRGAN~\cite{ledig2017photo} \hspace{-4mm} &
EDSR~\cite{lim2017enhanced} \hspace{-4mm}
\\
\includegraphics[width=0.24\textwidth]{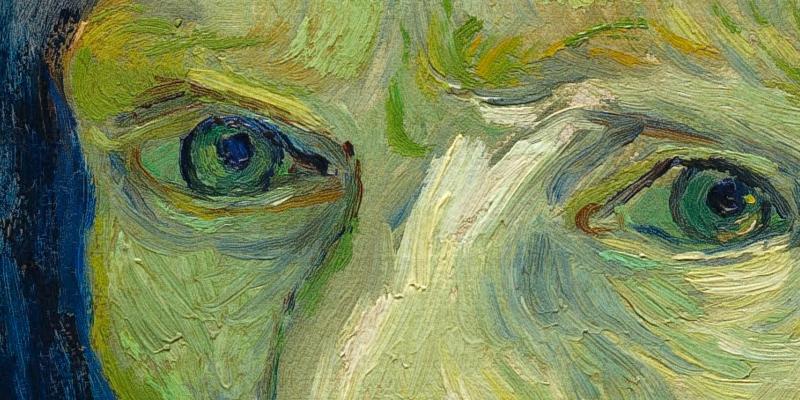} \hspace{-1mm} &
\includegraphics[width=0.24\textwidth]{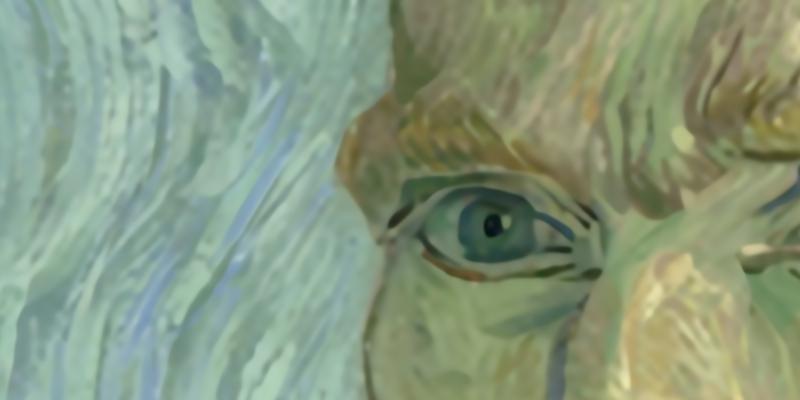} \hspace{-1mm} &
\includegraphics[width=0.24\textwidth]{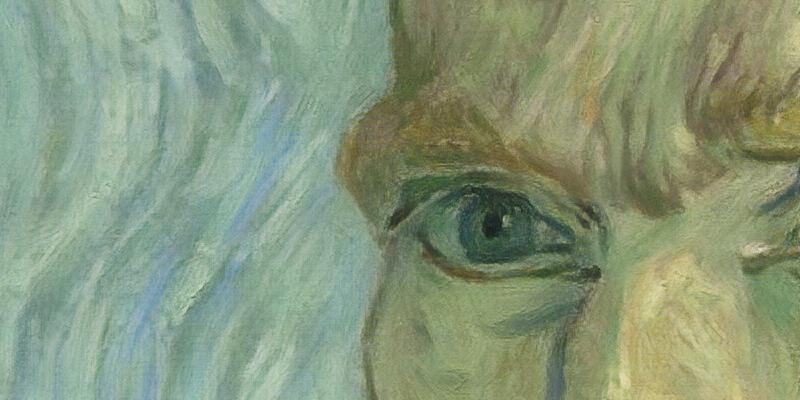} \hspace{-1mm} &
\includegraphics[width=0.24\textwidth]{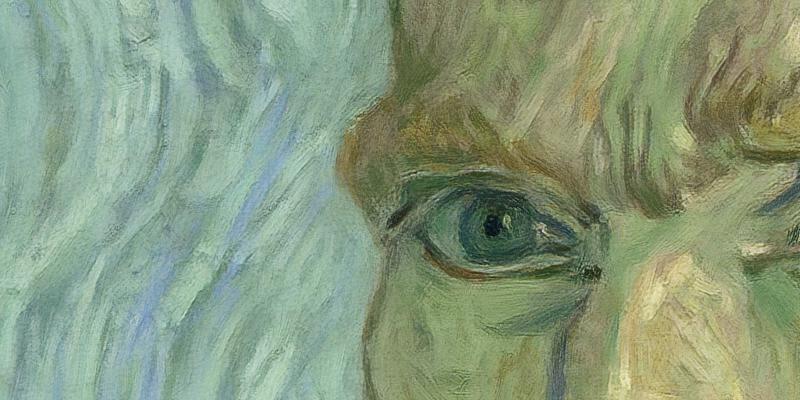} \hspace{-1mm}  
\\ 
Reference \hspace{-4mm} &
RCAN~\cite{zhang2018image} \hspace{-4mm} &
SRNTT~\cite{zhang2019image} \hspace{-4mm} &
Ours \hspace{-4mm}
\\
\end{tabular}
\end{adjustbox}

\end{tabular}
%\vspace{-4mm}
\caption{Visual results ($8\times$) of different SR methods on PaintHD}
\label{fig:visual_comp_x8}
%\vspace{-6mm}
\end{figure}

First, SISR methods would obtain higher PSNR and SSIM values than those of Ref-SR methods. This is reasonable because SISR methods mainly target to minimize MSE, which helps to pursue higher PSNR values. But, when the scaling factor goes to larger (\eg $16\times$), the gap among SISR methods also becomes very smaller. It means that it would be difficult to distinguish the performance between different SISR methods by considering PSNR/SSIM. 

Based on the observations and analyses above, we conclude that we should turn to other more visually-perceptual ways to evaluate the performance of SR methods, instead of only depending on PSNR/SSIM values. So, we further evaluate the PI values of each SR method. We can see `Ours-$\mathcal{L}_{rec}$' achieves lower PI values than those of `SRNTT-$\mathcal{L}_{rec}$', which is consistent with the analyses in Section~\ref{subsec:ablation_study}. SRNTT~\cite{zhang2019image} would achieve lower PI values than other SISR methods. It's predictable, as SRNTT transfers textures from high-quality reference. However, our method would achieve the lowest PI values among all the compared methods. Such quantitative results indicate that our method obtains outputs with better visual quality. To further support our analyses, we further conduct visual results comparisons and user study.

\begin{figure}[t]
%\newlength-4mm
%\setlength{-4mm}{-0.4cm}
\scriptsize
\centering
\begin{tabular}{cc}
% % one row
%\hspace{-0.4cm}
%\begin{adjustbox}{valign=t}
%\begin{tabular}{c}
%\includegraphics[width=0.2035\textwidth]{figs_jpg/ablation/pipeline/8x/ours_crop0.jpg}
%\\
%Kodak24: kodim24
%\end{tabular}
%\end{adjustbox}
%\hspace{-0.46cm}
\begin{adjustbox}{valign=t}
\begin{tabular}{cccccc}
\includegraphics[width=0.24\textwidth]{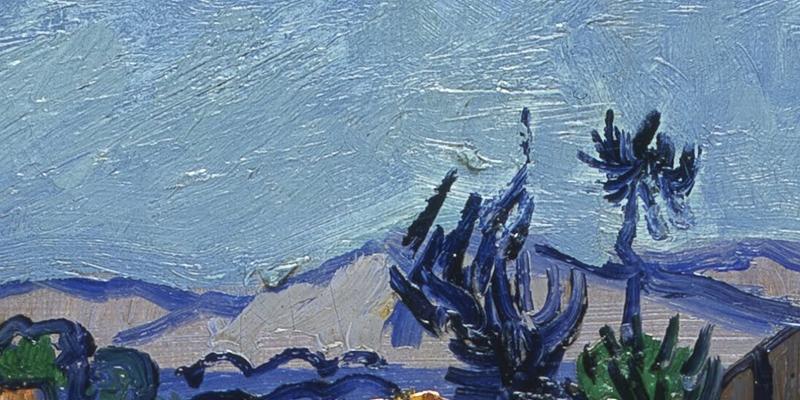} \hspace{-1mm} &
\includegraphics[width=0.24\textwidth]{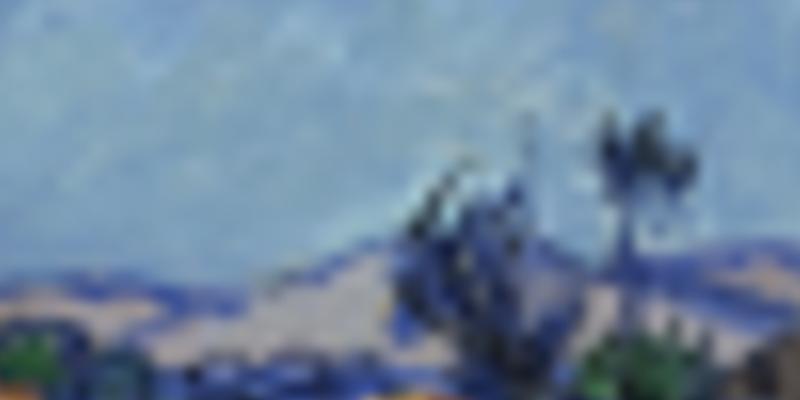} \hspace{-1mm} &
\includegraphics[width=0.24\textwidth]{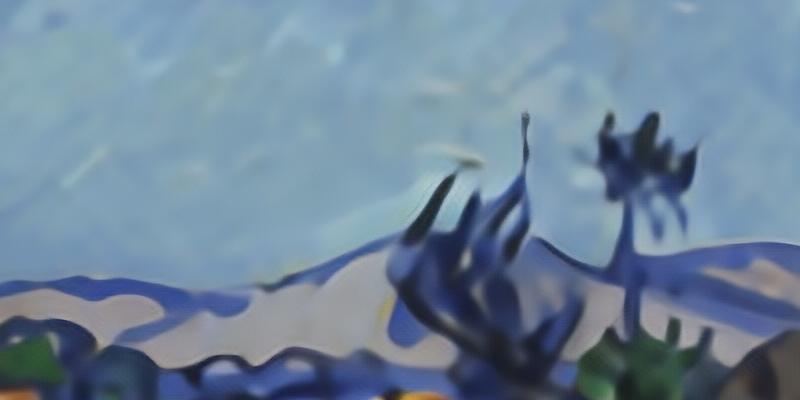} \hspace{-1mm} &
\includegraphics[width=0.24\textwidth]{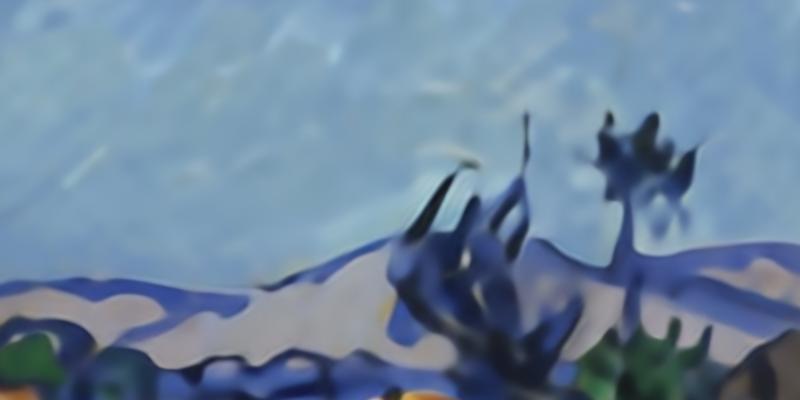} \hspace{-1mm} 
\\
HR \hspace{-4mm} &
Bicubic \hspace{-4mm} &
SRGAN~\cite{ledig2017photo} \hspace{-4mm} &
EDSR~\cite{lim2017enhanced} \hspace{-4mm}
\\
\includegraphics[width=0.24\textwidth]{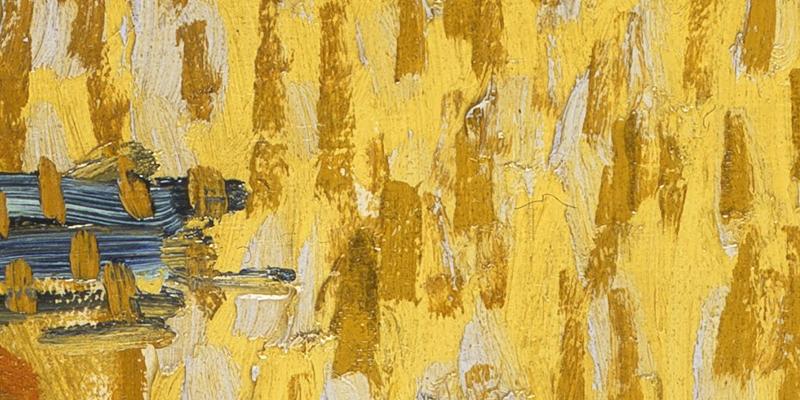} \hspace{-1mm} &
\includegraphics[width=0.24\textwidth]{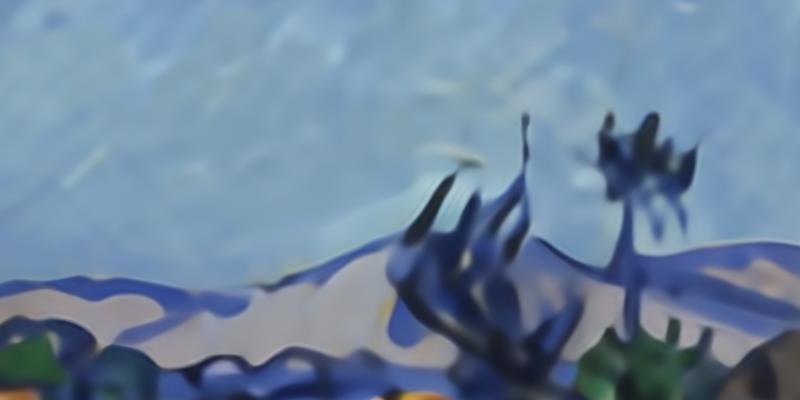} \hspace{-1mm} &
\includegraphics[width=0.24\textwidth]{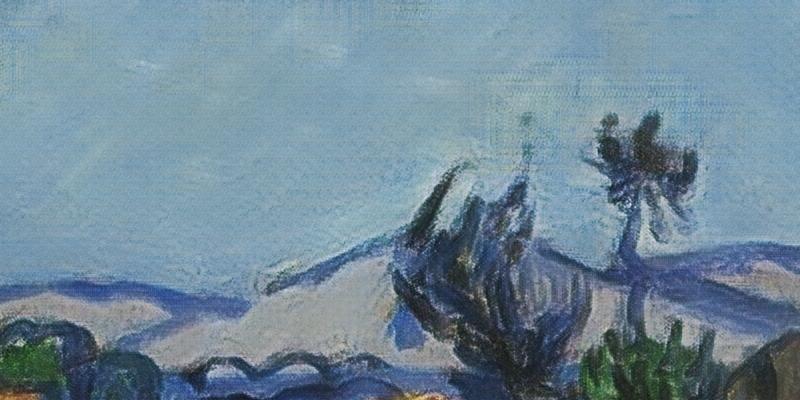} \hspace{-1mm} &
\includegraphics[width=0.24\textwidth]{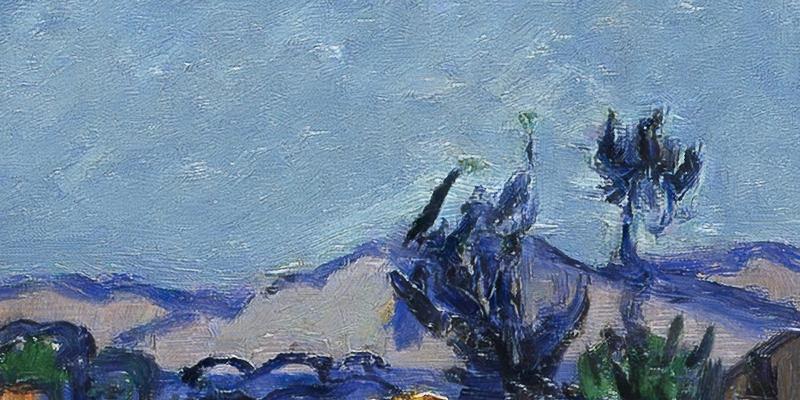} \hspace{-1mm}  
\\ 
Reference \hspace{-4mm} &
RCAN~\cite{zhang2018image} \hspace{-4mm} &
SRNTT~\cite{zhang2019image} \hspace{-4mm} &
Ours \hspace{-4mm}
\\
\end{tabular}
\end{adjustbox}

\\
\begin{adjustbox}{valign=t}
\begin{tabular}{cccccc}
\includegraphics[width=0.24\textwidth]{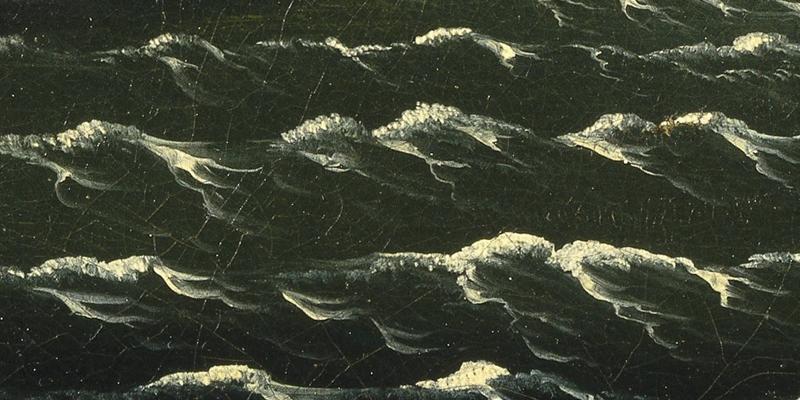} \hspace{-1mm} &
\includegraphics[width=0.24\textwidth]{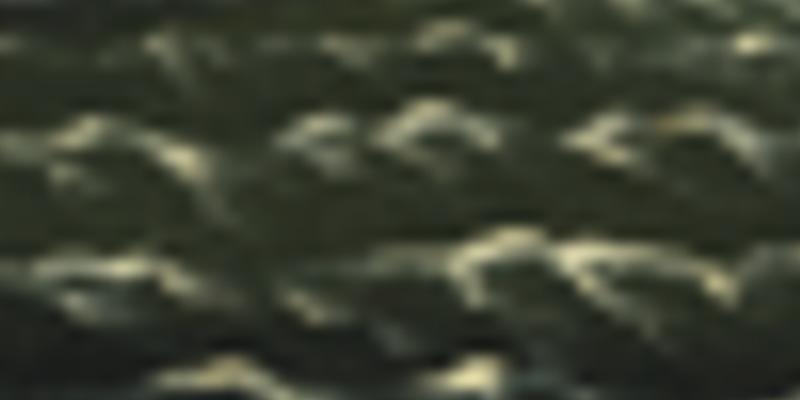} \hspace{-1mm} &
\includegraphics[width=0.24\textwidth]{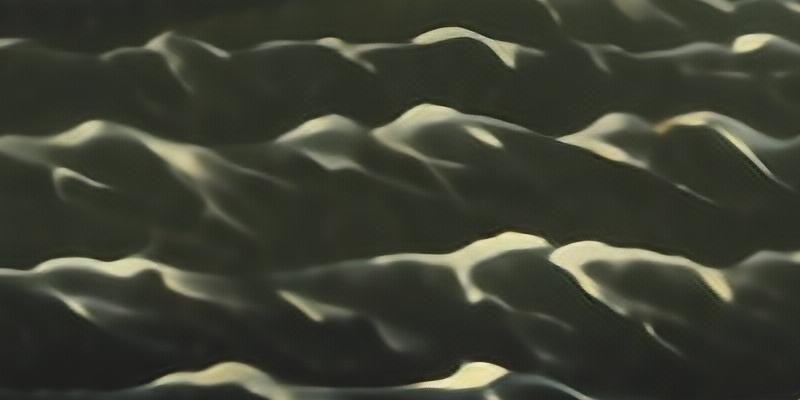} \hspace{-1mm} &
\includegraphics[width=0.24\textwidth]{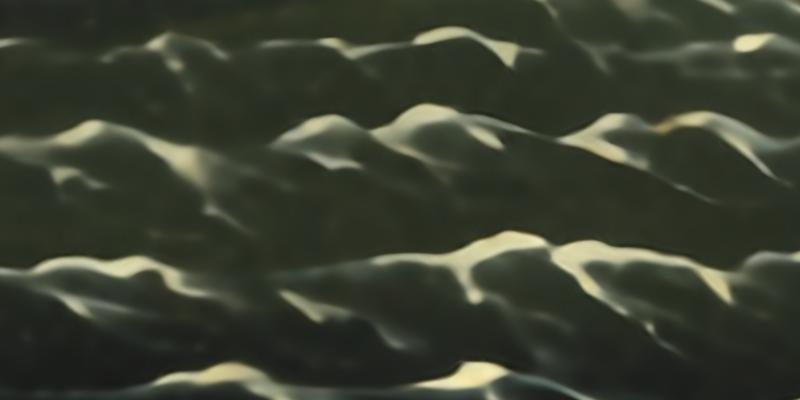} \hspace{-1mm} 
\\
HR \hspace{-4mm} &
Bicubic \hspace{-4mm} &
SRGAN~\cite{ledig2017photo} \hspace{-4mm} &
EDSR~\cite{lim2017enhanced} \hspace{-4mm}
\\
\includegraphics[width=0.24\textwidth]{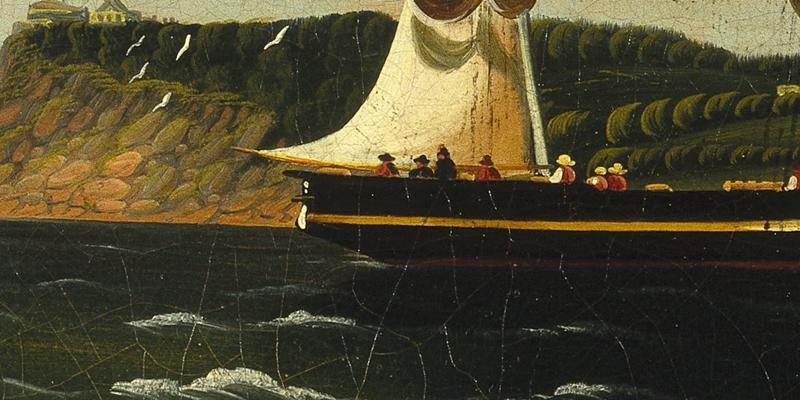} \hspace{-1mm} &
\includegraphics[width=0.24\textwidth]{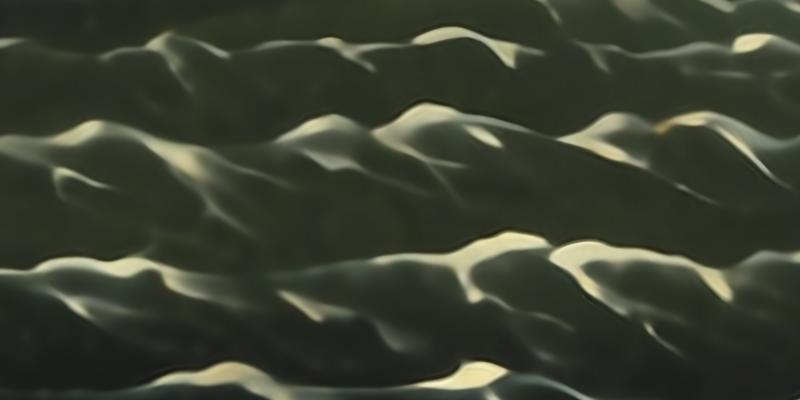} \hspace{-1mm} &
\includegraphics[width=0.24\textwidth]{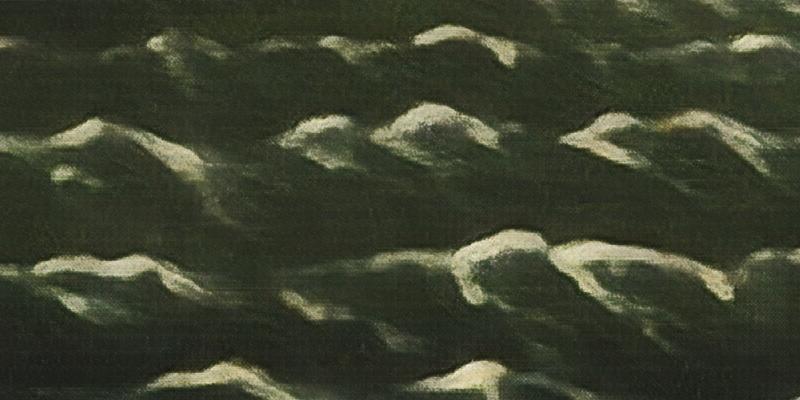} \hspace{-1mm} &
\includegraphics[width=0.24\textwidth]{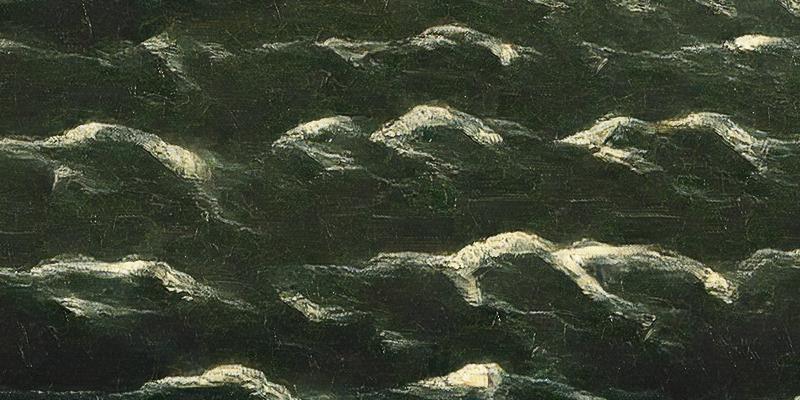} \hspace{-1mm}  
\\ 
Reference \hspace{-4mm} &
RCAN~\cite{zhang2018image} \hspace{-4mm} &
SRNTT~\cite{zhang2019image} \hspace{-4mm} &
Ours \hspace{-4mm}
\\
\end{tabular}
\end{adjustbox}

\end{tabular}
%\vspace{-4mm}
\caption{Visual results ($16\times$) of different SR methods on PaintHD}
\label{fig:visual_comp_x16}
%\vspace{-4mm}
\end{figure}

%\vspace{-4mm}
\subsection{Visual Comparisons}
\label{subsec:visual_comparisons}
%\vspace{-3mm}
As our PaintHD contains very high-resolution images with abundant textures, it's a practical way for us to show the zoom-in image patches for comparison. To better view the details of high-resolution image patches, it's hard for us to show image patches from too many methods. As a result, we only show visual comparison with state-of-the-art SISR and Ref-SR methods: SRGAN~\cite{ledig2017photo}, EDSR~\cite{lim2017enhanced}, RCAN~\cite{zhang2018image}, and SRNTT~\cite{zhang2019image}.   
	
We show visual comparisons in Figs.~\ref{fig:visual_comp_x8} and~\ref{fig:visual_comp_x16} for $8\times$ and $16\times$ cases respectively. Take $8\times$ case as an example, SISR methods could handle it to some degree, because the LR input has abundant details for reconstruction. But, SISR methods still suffer from some blurring artifacts due to use PSNR-oriented loss function (\eg $\ell_{1}$-norm loss). By transferring textures from reference and using other loss functions (\eg texture, perceptual, and adversarial losses), SRNTT~\cite{zhang2019image} performs visually better than RCAN. But SRNTT still can hardly transfer more detailed textures. In contrast, our method would obviously address the blurring artifacts and can transfer more vivid textures.
	
%When the case goes more challenging, namely $16\times$, both SISR methods and SRNTT would generate obvious over-smoothing artifacts (see Fig.~\ref{fig:visual_comp_x16}). The reasons for SISR methods are mainly that they aim to narrow the pixel-wise difference between the output $I_{SR}$ and the ground truth $I_{GT}$. Such an optimization way would encourage more low-frequency components, but restrain the generation of high-frequency ones. SRNTT is originally designed for $4\times$ upscaling, which restricts its ability for larger scaling factors. SRNTT neglects to pay more attention to the recovery of high-frequency components. In contrast, our method establishes trainable network from the original LR input to the target HR output firstly. Moreover, we focus on the reconstruction of high-frequency components more with wavelet texture loss. We even further relax the constraint between the output and ground truth by propose the degradation loss. As a result, our method alleviates the over-smoothing artifacts to some degree and recovers more detailed textures (see Fig.~\ref{fig:visual_comp_x16}). 

\begin{figure}[t]
\centering
\includegraphics[width=0.5\columnwidth]{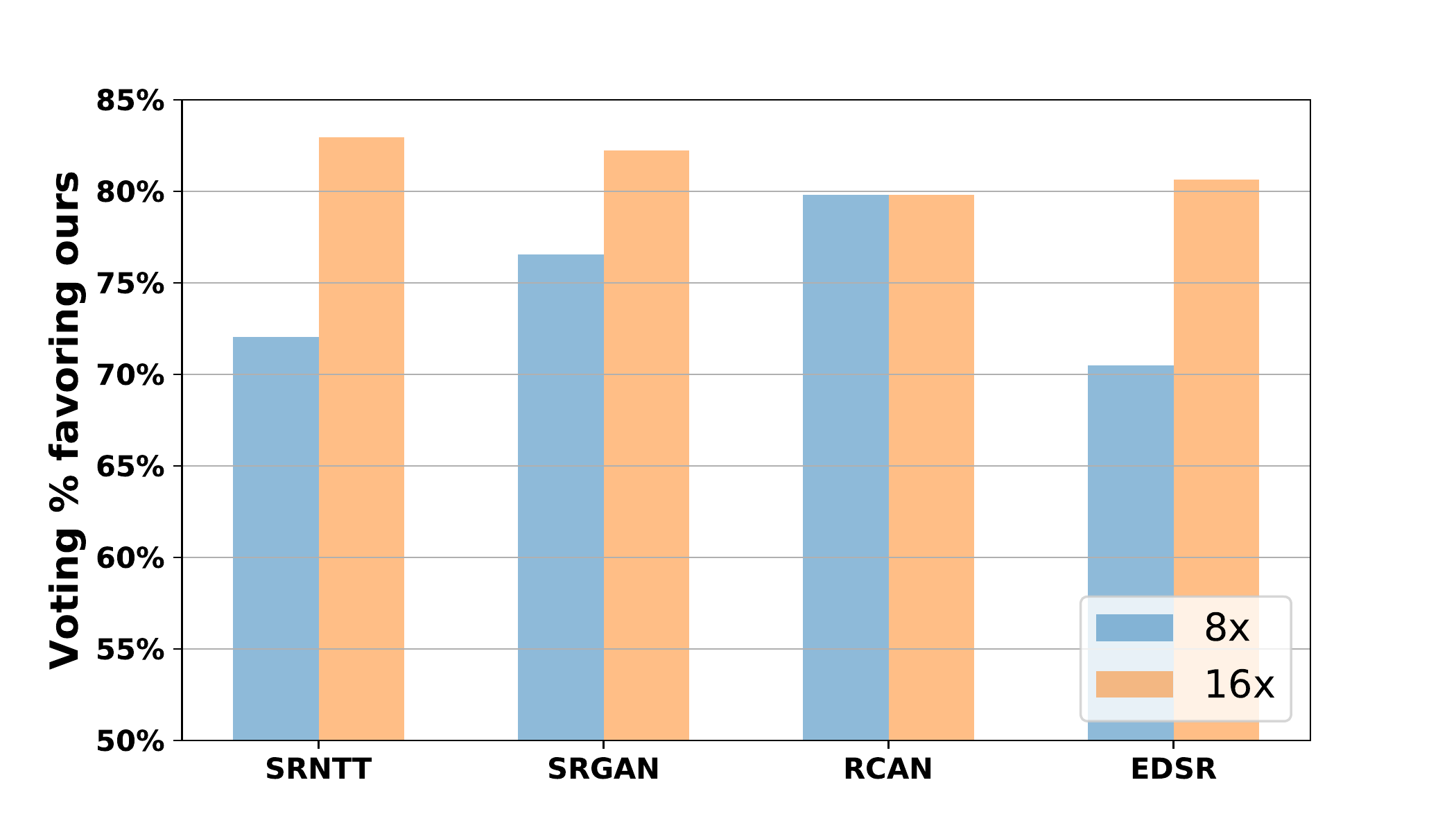}
%\vspace{-4mm}
\caption{User study on the results of SRNTT, SRGAN, RCAN, EDSR, and ours on the PaintHD and CUFED5 datasets. The bar corresponding to each method indicates the percentage favoring ours as compared to the method}
%The two colors indicate the different upscaling factors, \ie 8$\times$ and 16$\times$.
\label{fig:survey}
%\vspace{-5mm}
\end{figure}

%\vspace{-4mm}
\subsection{User Study}
\label{subsec:user_study}

Since the traditional metric PSNR and SSIM do not consistent to visual quality~\cite{johnson2016perceptual,ledig2017photo,sajjadi2017enhancenet,zhang2019image}, we conducted user study by following the setting of SRNTT~\cite{zhang2019image} to compare our results with those from other methods, \ie SRNTT~\cite{zhang2018densely}, SRGAN~\cite{ledig2017photo}, RCAN~\cite{zhang2018image}, and EDSR~\cite{lim2017enhanced}. The EDSR and RCAN achieve state-of-the-art performance in terms of PSNR/SSIM, while SRGAN (SISR) and SRNTT (Ref-SR) focus more on visual quality. All methods are tested on a random subset of CUFED5 and PaintHD at the upscaling factor of 8$\times$ and 16$\times$. In each query, the user is asked to select the visually better one between two side-by-side images super-resolved from the same LR input, \ie one from ours and the other from another method. In total, we collected 3,480 votes, and the results are shown in Fig.~\ref{fig:survey}. The height of a bar indicates the percentage of users who favor our results as compared to those from a corresponding method. In general, our results achieve better visual quality at both upscaling scales, and the relative quality at 16$\times$ further outperforms the others. 
%The main reason lies in the texture transfer from references. With the increase of the upscaling factor, more details are lost in the LR input, which is difficult to be recovered solely by the deep model. Thus, externally high-frequency information tends to be more important to texture recovery, which causes the gap from 5\% to 10\% between the results of 8$\times$ and 16$\times$.

%\subsubsection{Vote Best from All.}

% \begin{table}[htbp]
% \scriptsize
% %\footnotesize
% %\small
% %\normalsize
% \centering
% \begin{center}
% %\caption{Ablation investigation of contiguous memory (CM), local residual learning (LRL), and global feature fusion (GFF). We observe the best performance (PSNR) on Set5 with scaling factor $\times2$ in 200 epochs.} 
% %\vspace{-1mm}
% \caption{Percentage (\%) of the votes that each method received.}
% \label{tab:results_user_study} 
% %\vspace{-3mm}
% %\begin{tabular*}{82.4mm}{@{\extracolsep{-0.75mm}}|c|c|c|c|c|c|c|c|c|}
% \begin{tabular}{|c|c|c|c|c|c|c|c|c|c|c|c|}
% \hline
% Method & SRGAN  & EDSR  & RCAN & SRNTT & Ours 
% \\
% \hline
% 8$\times$& 21.41  & 11.31  & 12.67 & 11.55 & \textbf{33.45} 
% \\
% \hline
% 16$\times$& 21.41  & 11.31  & 12.67 & 11.55 & \textbf{33.45} 
% \\
% \hline
% \end{tabular}
% \end{center}
% %%\vspace{-6mm}
% \end{table}
%\vspace{-4mm}
\subsection{Effect of Different References}
\begin{figure}[t]
%\newlength-4mm
%\setlength{-4mm}{-0.4cm}
\scriptsize
\centering
\begin{tabular}{cc}
% % one row
%\hspace{-0.4cm}
%\begin{adjustbox}{valign=t}
%\begin{tabular}{c}
%\includegraphics[width=0.2035\textwidth]{figs_jpg/ablation/pipeline/8x/ours_crop0.jpg}
%\\
%Kodak24: kodim24
%\end{tabular}
%\end{adjustbox}
%\hspace{-0.46cm}
\begin{adjustbox}{valign=t}
\begin{tabular}{cccccc}
\includegraphics[width=0.24\textwidth]{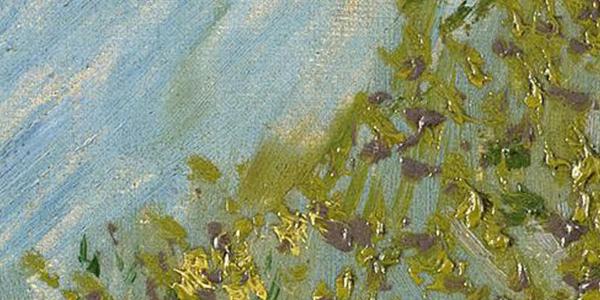} \hspace{-1mm} &
\includegraphics[width=0.24\textwidth]{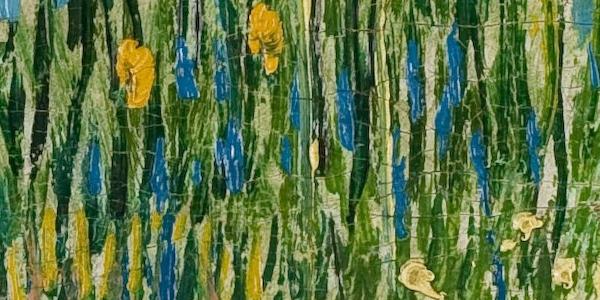} \hspace{-1mm} &
\includegraphics[width=0.24\textwidth]{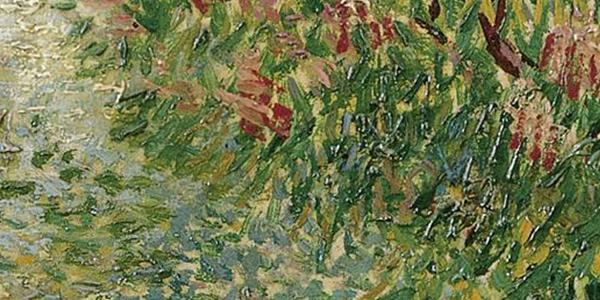} \hspace{-1mm} &
\includegraphics[width=0.24\textwidth]{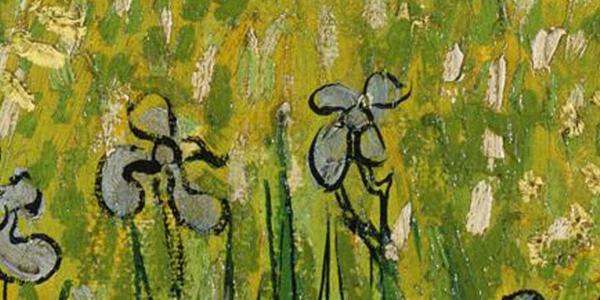} \hspace{-1mm} 
\\
HR \hspace{-4mm} &
Reference 1 \hspace{-4mm} &
Reference 2 \hspace{-4mm} &
Reference 3 \hspace{-4mm}
\\
\includegraphics[width=0.24\textwidth]{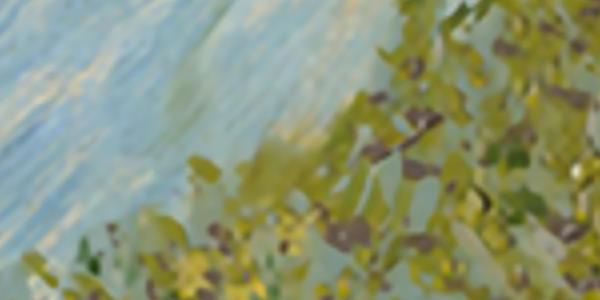} \hspace{-1mm} &
\includegraphics[width=0.24\textwidth]{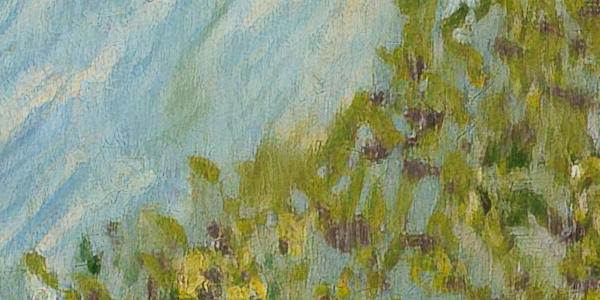} \hspace{-1mm} &
\includegraphics[width=0.24\textwidth]{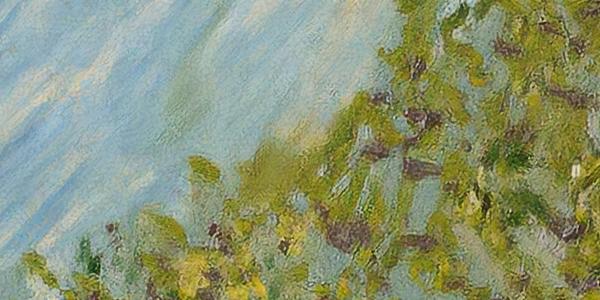} \hspace{-1mm} &
\includegraphics[width=0.24\textwidth]{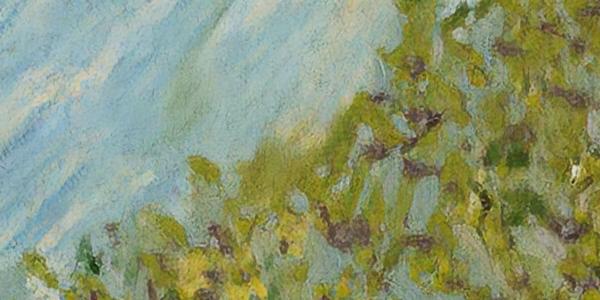} \hspace{-1mm}  
\\ 
Bicubic \hspace{-4mm} &
Result 1 (Ours) \hspace{-4mm} &
Result 2 (Ours) \hspace{-4mm} &
Result 3 (Ours) \hspace{-4mm}
\\
\end{tabular}
\end{adjustbox}

\end{tabular}
%\vspace{-4mm}
\caption{Visual results with scaling factor 8$\times$ using different reference images}
\label{fig:visual_diff_ref}
%\vspace{-6mm}
\end{figure}

%\vspace{-3mm}
For Ref-SR methods, investigation on the effect from references is an interesting and opening problem, \eg how the references affect SR results, how to control (\ie utilize or suppress) such effect, etc. This section intends to explore the effect of references in the proposed Ref-SR method. As shown in Fig.~\ref{fig:visual_diff_ref}, the same LR input is super-resolved using different reference images, respectively. We can see that the results keep similar content structures as the input. If we give a further look at the details, we find each result has specific textures from the corresponding reference. It indicates that our method keeps the main structures to the LR input, but also adaptively transfers texture details from reference.

%\vspace{-4mm}
\section{Conclusions}
%\vspace{-3mm}
We aim to hallucinate painting images with very large upscaling factors and transfer high-resolution (HR) detailed textures from HR reference images. Such a task could be very challenging. The popular single image super-resolution (SISR) could hardly transfer textures from reference images. On the other hand, reference-based SR (Ref-SR) could transfer textures to some degree, but could hardly handle very large scaling factors. We address this problem by first construct an efficient Ref-SR network, being suitable for very large scaling factor. To transfer more detailed textures, we propose a wavelet texture loss to focus on more high-frequency components. To alleviate the potential over-smoothing artifacts caused by reconstruction constraint, we further relax it by proposed a degradation loss. We collect high-quality painting dataset PaintHD, where we conduct extensive experiments and compare with other state-of-the-art methods. We achieved significantly improvements over both SISR and Ref-SR methods. 
%We believe such a Ref-SR network has promising benefits to general natural images. 

\noindent\textbf{Acknowledgments}. This work was supported by the Adobe gift fund.

\clearpage
\bibliographystyle{splncs04}
\bibliography{SR_conf_bib}
\end{document}